# Disorder protects collagen networks from fracture

F. Burla[1] †, S. Dussi[2] †, C. Martinez-Torres[1,3], J. Tauber[2], J. van der Gucht[2*], G.H. Koenderink[1,3*]

*[1] AMOLF, Department of Living Matter, Biological Soft Matter group, Science Park 104, 1098 XG Amsterdam, the Netherlands*

*[2] Physical Chemistry and Soft Matter, Wageningen University & Research, Stippeneng 4, 6708 WE Wageningen, Netherlands.*

*[3] Department of Bionanoscience, Kavli Institute of Nanoscience Delft, Delft University of Technology, Van der Maasweg 9, 2629 HZ Delft, the Netherlands*

*† These authors contributed equally to this work.*

** Corresponding authors.*

*Email addresses: g.h.koenderink@tudelft.nl, jasper.vandergucht@wur.nl*

## Abstract

Collagen forms the structural scaffold of connective tissues in all mammals. Tissues are remarkably resistant against mechanical deformations because collagen molecules hierarchically self-assemble in fibrous networks that stiffen with increasing strain. Nevertheless, collagen networks do fracture when tissues are overloaded or subject to pathological conditions such as aneurysms. Prior studies of the role of collagen in tissue fracture have mainly focused on tendons, which contain highly aligned bundles of collagen. By contrast, little is known about fracture of the orientationally more disordered collagen networks present in many other tissues such as skin and cartilage. Here, we combine shear rheology of reconstituted collagen networks with computer simulations to investigate the primary determinants of fracture in disordered collagen networks. We show that the fracture strain is controlled by the coordination number of the network junctions, with less connected networks fracturing at larger strains. The hierarchical structure of collagen fine-tunes the fracture strain by providing structural plasticity at the network and fiber level. Our findings imply that structural disorder provides a protective mechanism against network fracture that can optimize the strength of biological tissues.

## Introduction

Collagen is the most abundant protein in the human body. Secreted in the extracellular space by cells, collagen assembles in networks of thick rope-like fibrils that shape and reinforce tissues and provide a scaffold for cell growth and movement[1]. The spatial organization of these networks varies widely among tissues, from aligned bundles in tendons to randomly oriented (isotropic) networks in skin. Isotropic collagen networks tend to have a low connectivity, because the fibrils are mainly joined in three-fold junctions by branches and in four-fold junctions by crosslinks. As a result, the network connectivity is below the Maxwell criterion of six required for mechanical stability of random networks of springs[2,3]. It was recently shown that the subisostatic architecture offers a great mechanical advantage for collagen networks, because it causes them to be soft at small deformations, primarily due to fibril bending, yet stiff at large deformations, due to fibril alignment and a corresponding transition from fibril bending to stretching[3–5]. Nevertheless, tissues still fracture when exposed to large deformations, especially in pathological conditions such as injuries[6], surgical interventions[7], aneurysms[8,9] and hydraulic fracture of tumors[10,11].

Fracture of collagen has so far mainly been investigated in tendons[12,13], where collagen fibrils organize in long cable-like structures optimized to withstand large axial loads [14,15]. Because of the unidirectional fiber orientations, the fracture of tendons is mainly governed by molecular properties of the fibrils, which vary among functionally distinct tendons[13] and change upon age-related enzymatic crosslinking reactions[16,17] and during diseases[18]. However, collagen in many other connective tissues assembles in disordered networks that lack a preferential orientation. Examples are skin[19], cartilage[20], vitreous humour[21], and the aortae[22]. Interestingly, research on aortae fracture in the context of aneurysms[9] as well as studies using tissue models[23–26] have revealed that isotropic tissues fracture at higher strain than aligned tissues like tendon, because the propagation of cracks is contained. This observation suggests that isotropic networks might be optimized to withstand larger strains. However, a mechanistic understanding of the role of network architecture in tissue fracture has so far been lacking, due to the complexity of living tissues.

Here, we investigate the mechanisms that protect isotropic collagen networks from fracture by performing quantitative measurements of shear-induced fracture of reconstituted collagen networks, both experimentally and computationally. Experimentally, we control the collagen structure from the network level (mesh-size and connectivity) down to the fiber level (diameter and intrafibrillar crosslinking) by reconstituting networks of collagen purified from different animal and tissue sources and by exploiting the known sensitivity of collagen self-assembly to the polymerization temperature[27,28]. By comparing our results against a computational model of network fracture, we find that the connectivity of the network, defined as the mean number of fibers meeting at a junction, is the main determinant of collagen fracture. We can explain almost all of our findings by using a threshold strain for fiber fracture in the range of 10-20%, a value that is consistent with prior tensile tests on single collagen fibrils[29,30]. Furthermore, the computational model enables us to assess the contributions of system size and detailed network and fiber properties on the fracture behavior. Molecular effects, such as intrafibrillar crosslinking, and network properties, such as branching, modulate the fracture strain by setting the degree of plasticity. Our results are important not only for understanding how disorder protects collagen networks -and therefore, living tissues- from fracture, but also for the rational design of synthetic fibrillar materials resistant to strain-induced breakage.

### Results

To test the mechanical resistance of collagen networks against fracture, we perform rheology experiments on reconstituted collagen networks polymerized between the plates of a custom-built confocal rheometer. The bottom plate of the rheometer is stationary and optically transparent to allow direct visualization of changes in network structure in response to mechanical deformation using an inverted confocal microscope (Figure 1a). In order to assess collagen fracture, we apply a linear strain ramp $\gamma$ on the networks by rotating the top plate and measuring the resulting shear stress $\sigma$. As illustrated in Figure 1b, the stress initially increases linearly with strain, as expected for the linear elastic regime. However, above a threshold strain, the stress starts to deviate from this linear behavior and shows an upturn indicative of network stiffening. The stress eventually reaches a maximum value, which we call the peak stress $\sigma_p$, associated to a peak strain $\gamma_p$. Beyond the stress peak, the shear stress decreases, which is symptomatic of fracture. Images taken during the strain ramp at a fixed height of 20 μm above the bottom plate of the rheometer indeed reveal network fracture, as signaled by the onset of fibril motion, the breaking of connections and a decrease in fluorescence intensity in the imaging plane (Fig. 1c and Supplementary Video S1). The eventual disappearance of the network from the field of view suggests that a crack appears in the sample. However, fracture is always first observed at strains beyond $\gamma_p$ (see Supplementary Figs. S1-S2). This may be partly explained from the fact that the macroscopic strain we report here corresponds to the strain at the edge of the sample, while our imaging area is located at a radial distance halfway from the center so the local strain is only 50% of the macroscopic strain. Additionally, it is possible that cracks first form in areas outside the field of view and are observable only once they propagate to the field of view. In some cases, the plane of fracture is localized above the imaging plane and we cannot observe network fracture at all. Post-fracture imaging of the samples over an extended height range confirms that fracture always occurred within the bulk of the network rather than at the bottom rheometer plate (Supplementary Fig. S3).

In order to understand which structural parameters are predictive of collagen network fracture, we prepare networks with a wide range of architectures by polymerizing collagen extracted from different animal and tissue sources at different temperatures[31] (Figure 2a). We are thus able to control the structure both at the network level (mesh size and coordination number) and at the single-fiber level (diameter) (see Figure 2b and quantification in Supplementary Fig. S4). Furthermore, we vary the fibrils' properties by comparing collagen molecules with and without telopeptide end sequences, the disordered extensions of the collagen triple helix that mediate intrafibrillar crosslinking[32]. When we subject these networks to the strain ramp protocol, we measure peak strains $\gamma_p$ that vary over a remarkably large range, from 20% all the way up to nearly 90% (see Figure 2c). The peak strains are independent of strain rate and plate diameter (Supplementary Figs. S5-S6), again showing that the networks fail cohesively and not at the interface with the rheometer plate. We also note that changes in the surface chemistry of the plates do not significantly influence the value of the peak strain compared to its overall variation (Supplementary Figs. S7).

We do find a strong correlation between the fracture strain $\gamma_p$ and the critical strain $\gamma_c$, where the networks undergo the transition from the soft bend-dominated regime to the stiff stretch-dominated regime, as shown in the inset of Figure 3a. This correlation hints at the possibility that both strains are controlled by the collagen networks' average connectivity[5,29,31]. By mapping the nonlinear elastic response, and in particular the critical strain measured by shear rheology, onto computational predictions for fibrillar networks, we can extract the average network connectivity $\langle z \rangle$. This mapping method was recently validated for collagen networks over a wide range of connectivities[31]. We find that our collagen networks have $\langle z \rangle$ ranging between 2.9 and 3.7 (see Supplementary Fig. S8). We observe from Figure 3a that the fracture strain monotonically decreases from ~90% for $\langle z \rangle=2.9$ to ~20% for $\langle z \rangle=3.7$ (see also Supplementary Fig. S9). By contrast, we observe no correlation of the fracture strain with the parameters characterizing the fibers themselves, such as the fibers' diameter and bending rigidity, nor with the network mesh size (Supplementary Fig. S10). Our data therefore strongly suggest that the network connectivity is the dominant factor in setting the fracture strain.

This observation is consistent with recent simulations and experiments on other disordered systems, including elastic spring networks and metamaterials, whose fracture is governed by connectivity[33–36]. We therefore compare our results to fracture simulations of coarse-grained fiber networks composed of $L$ by $L$ nodes, where each bond is modelled as an elastic spring that fractures irreversibly when its axial deformation exceeds a rupture threshold $\lambda$. When we apply a shear deformation to the simulated networks, bonds align along the direction of the deformation and the network strain-stiffens (Fig. 3b), consistently with earlier findings[3,5]. In the strain-stiffening regime, the stress is heterogeneously distributed, concentrating in regions of aligned load-bearing bonds referred to as force chains. The lower the average network connectivity, the more heterogeneous is the stress distribution (see Supplementary Fig. S11). Above a threshold shear strain, bonds first start fracturing in an uncorrelated fashion and some force chains disappear, while new ones appear. When the strain reaches $\gamma_p$, an individual fracturing event triggers the formation of a large crack (see middle and right snapshots in Fig. 3b). Although this crack has not yet propagated through the entire sample, it does cause a large and abrupt stress drop. After the peak, the stress does not completely vanish and further small stress drops are observed, indicative of the breakage of the remaining few force chains.

To test whether this simple model can account for the connectivity-dependent fracture of collagen network, we perform a series of simulations for networks with different <z>-values spanning the experimental range and assuming a rupture threshold $\lambda$ of either 10 or 20%, bracketing the range observed in single-fibril rupture experiments[29,30]. As shown in Fig. 3a, the simulations are in first instance consistent with the experimentally measured dependence of the fracture strain on connectivity. Although collagen fibers have a finite bending rigidity, additional simulations show that fiber bending plays a negligible role, as fracture always occurs at large enough strains such that the networks are past the transition from the bending- to the stretching-dominated regime (see Supplementary Fig. S12). We note that the post-peak mechanical response in simulations and experiments is different: the simulated stress-strain curves show an abrupt post-peak behavior indicative of brittle fracture (Figure 3b), while the experimental stress-strain curves exhibit a more gentle, ductile fracture (Supplementary Fig. S9). The most likely origin of this difference lies in the different system sizes in experiments and simulations. To test this idea, we first verify that simulations of sheared networks reveal more brittle fracture with increasing system size (Supplementary Figures S13-S14), in agreement with our recent simulations of networks under uniaxial extension[36]. We then perform fracture experiments for varying gap sizes and again observe a size-dependent change in ductility, even though the peak strain shows no gap size dependence for gaps larger than 200 μm (Supplementary Figures S15-S16), consistent with a previous study[37]. Finally, we perform simulations keeping the vertical dimension $L_y$ fixed while increasing the lateral size $L_x$, in order to mimic the experimental conditions where the lateral dimensions (4 cm) are almost two orders of magnitude larger than the gap size. We find that the post-peak stress-strain response is indeed smoother with increasing $L_x$ (Supplementary Figures S15-S16), capturing the experimental response.

Interestingly, we notice that for low connectivity (between 2.9 and 3.0) the experimentally measured peak strain is systematically higher than the computational predictions (see also Supplementary Figure S17). This discrepancy might be due to structural plasticity in the collagen networks due to their hierarchical structure, which we have not yet considered in our computational model. Plastic mechanisms can occur either at the network level, with opening up of branches, or at the fibril level, where fibril lengthening can occur via monomer sliding[38–40] (see Figure 4a). We expect sliding to be particularly prominent in case of the non-crosslinked fibrils formed from collagen molecules that lack telopeptides. To experimentally test the role of fibril plasticity, we perform cyclic strain ramp tests (with a maximum strain equal to $\gamma_p/2$ to prevent fracture), comparing two bovine collagen molecules that give networks with a similar connectivity (<z>~3.4) but where one collagen lacks telopeptides ('atelo'), while the other has intact telopeptides ('telo'). As shown in Fig. 5b, we observe a larger hysteresis between the loading and unloading curves for the (non-crosslinked) atelocollagen compared to the (crosslinked) telocollagen. We also observe that the atelocollagen is softer than the telocollagen and exhibits a larger softening interval before fracture, consistent with a higher degree of plasticity (see Supplementary Figure S18). To test for the role of plasticity at the network level, we compare two collagen networks that are both crosslinked via telopeptides but have different connectivities: telocollagen from bovine dermal skin (<z>=3.42) and rat tail collagen (<z>=2.9). In this case a larger hysteresis between the loading and unloading curves is observed for the <z>=2.9 (rat tail) collagen as compared to the <z>=3.42 (bovine) collagen, consistent with our hypothesis that branches can accommodate larger strains than crosslinked (z=4) junctions.

To obtain more insight in the influence of branch and fiber plasticity on network fracture, we finally consider the shape of the stress-strain curves. We compare simulations and experiments by normalizing the curves with the peak strain $\gamma_p$ and stress $\sigma_p$ (Fig. 4c). After inspecting the influence of all the simulation parameters (network connectivity, fiber rupture threshold, system size, fiber bending) on the shape of the curve (see Supplementary Figs. S19-S20), we choose the interval $0.6 < \gamma/\gamma_p < 0.9$ to quantify the difference between the curves and identify the simulation parameters that best capture the experimental curves. For crosslinked (telocollagen) networks at high connectivity, we find that the experiments agree well with network simulations that do not include any plastic effects (Fig. 4c (top)). By contrast, the stress-strain curves of low-connectivity and uncrosslinked (atelocollagen) networks can only be mapped onto the simulations if we allow for bond lengthening before fracture. We implement bond lengthening in the simulations by allowing the elastic springs to irreversibly extend above a lengthening threshold $l_m$. For simplicity, we assume that when the spring length reaches $l_m$, the rest length permanently changes from its initial value $r_1=1$ to a value $r_2 > r_1$. As before, we assume that the springs furthermore irreversibly break as soon as their deformation exceeds $\lambda$ (i.e. when their length is $r_1+\lambda$), as shown in Fig. 4a. For comparison with the experiments, we scan different combinations of $l_m$ and $r_2$ and calculate the difference between the simulated and experimental stress-strain curves (after normalization) and peak strains $\gamma_p$ (Supplementary Fig. S21). As shown in Fig. 4c (middle), we obtain a good match between measurements on (bovine) atelocollagen networks and simulations only for $l_m \sim 1.10$. An onset strain of 10% for fiber lengthening is consistent with previous tensile measurements on isolated collagen fibrils[41–43]. By contrast, for the telocollagen (rat tail) networks (Rat37, $\langle z\rangle=2.9$) we only find reasonable agreement with simulations for values of $l_m \sim 1.03$ (Fig. 4c, bottom and Supplementary Fig. S21). This small value of $l_m$ suggests that network-level plasticity, rather than fiber lengthening, modulates the fracture of these loosely connected ($\langle z\rangle\sim3$) networks.

The two parameters $l_m$ and $r_2$ together control the onset and degree of plastic effects due to bond lengthening in the simulations. As shown in Fig. 4d (where we fixed the rupture threshold $\lambda=30\%$), bond lengthening only enhances the peak strain $\gamma_p$ (blue up-pointing triangles) compared to the purely elastic limit in case of small $l_m$. By contrast, bond lengthening causes networks with larger $l_m$ to fracture earlier (red down-pointing triangles) compared to the elastic limit. Apparently, lengthening effects are not always beneficial for delaying fracture. This can be rationalized with the fact that the stress distribution is highly inhomogeneous for networks with low connectivity, where very few bonds are highly stressed (Supplementary Fig. 11). Lengthening is beneficial for alleviating stress concentration in these few bonds only if lengthening already occurs at small deformation (small $l_m$). If the lengthening occurs at large $l_m$, the already high stress carried by the few bonds will not be efficiently redistributed in the network due to its low connectivity. As a consequence, these few bonds will lengthen without releasing enough stress to their neighbors so they eventually fracture. The effective rupture threshold of these bonds is therefore determined by $l_m$ and is lower than their intrinsic threshold $\lambda$, so the entire network will fracture earlier than in the elastic limit. Future studies should employ less coarse-grained models, able to describe the mechanical response of bundles and branches, to further unveil the microscopic origin of these toughening mechanisms.

## Conclusions

Collagen forms the structural scaffold of living tissues, where it is often present in the form of a disordered network of fibrils with random orientations. Large mechanical stress and pathological situations such as aneurysms can threaten the mechanical integrity of collagenous tissues. It is known that fracture of tendons, which feature a strongly anisotropic collagen structure, is governed by the molecular makeup of collagen fibers. Here, we show that fracture of isotropic collagen networks representative of tissues such as skin and cartilage is instead governed by the average connectivity $\langle z\rangle$. We find that orientational disorder combined with subisostatic connectivity protect collagen networks against fracture. This correlation between network fracture strain and connectivity is similar to that found in simpler elastic networks, such as spring networks and metamaterials. However, the hierarchical complexity of collagen does provide additional protection of collagen networks against fracture by introducing mechanisms such as branch opening or fiber lengthening that accommodate additional strain. Our computational model provides a minimal yet powerful way to predict the fracture strain of collagenous tissues and other fibrillar materials from first principles. Our findings provide new routes to design strong fibrous materials based on the combination of a random subisostatic architecture with controlled plasticity at the network or fibril scale[44].

### Acknowledgments

We gratefully acknowledge F.C. MacKintosh (Rice University, Texas) for extensive scientific discussions and M. Dompè (WUR) for a critical reading of the manuscript. We furthermore thank M. van Hecke (AMOLF/Leiden University) for kindly lending us the confocal rheometer head and for critically reading the manuscript; M. Verweij, J.B. Aans and D.-J. Spaanderman (AMOLF) for help with building the confocal rheometer; and M. Vinkenoog (AMOLF) for help with CNA35 protein purification. The work of F.B. and G.H.K. is part of the Industrial Partnership Programme Hybrid Soft Materials that is carried out under an agreement between Unilever Research and Development B.V. and the Netherlands Organisation for Scientific Research (NWO). The work of S.D, J.T. and J.v.d.G. is part of the SOFTBREAK project funded by the European Research Council (ERC Consolidator Grant).

**Author contributions**

F.B. performed research (experiments) supervised by G.H.K. S.D. performed research (simulations) supervised by J.v.d.G. C.M-T. contributed to the acquisition of experimental images. J.T. contributed computational tools. F.B. and S.D. analyzed data. All authors contributed to the interpretation of the results and to the writing of the manuscript. F.B. and S.D. contributed equally to this work.

## Material and methods

### Sample preparation

Collagen networks were reconstituted from commercially available collagen purified from different animal species (cow, rat, human) and from different tissues (dermis, tendon) (see Supplementary Table 1). Moreover, we compared collagens obtained by two different extraction methods: telocollagen obtained by acid solubilization, which has intact telopeptide end sequences necessary for interfibrillar crosslinking, and atelocollagen obtained by pepsin solubilization, which lacks these telopeptides [27-28]. Finally, we also compared two different fibril-forming collagen types: type I, which is ubiquitous in all tissues except cartilage, and type II, which is characteristic of cartilage[45]. All samples were prepared on ice, to prevent early collagen polymerization, by first weighing collagen in an Eppendorf tube and subsequently adding water, 10x-concentrated PBS (Phosphate Buffered Saline, NaCl 0.138 M; KCl - 0.0027 M, pH 7.4, Sigma Aldrich), and an adequate amount of 0.1 M NaOH (sodium hydroxide, Sigma Aldrich) in order to obtain the final desired concentration of collagen (0.5 or 1 mg/mL) in a solution of PBS and at a pH of 7.4. Each sample was subsequently vortexed for a few seconds and then quickly placed in the measurement cell (cone-plate geometry or plate-plate geometry for rheometry, glass flow cell for microscopy, Eppendorf tube for electron microscopy) to start polymerization at the desired temperature. For the confocal rheometer experiments, we used a fluorescently tagged bacterial protein (CNA35-eGFP) that specifically binds to collagen fibrils[46] at a CNA/collagen molar ratio of 1:10.

### Rheological measurements

Strain-stiffening and fracture experiments were performed using an Anton Paar Physica MCR501 rheometer, with a stainless steel, cone-plate geometry where the cone had a 40 mm diameter and 1° cone angle. We verified the absence of wall slippage by repeating experiments with steel cone-plate cells with diameters of 20 and 30 mm. To test for gap size effects, we used plate-plate geometries with a gap of 100 $\mu$m, 250 $\mu$m, 500 $\mu$m or 750 $\mu$m and a diameter of 40 mm. Collagen solutions were allowed to polymerize for two hours between the rheometer plates at a constant temperature (27, 30, 34 or 37°C) maintained by a Peltier plate. During polymerization, the evolution of the linear shear moduli was monitored with a small amplitude oscillatory strain (0.5% strain, 0.5 Hz) protocol. The steady-state values of the linear viscoelastic moduli reached after two hours were calculated as an average over the last ten data points of the polymerization curve. We report averages of at least 3 independent measurements while the error bars represent the standard error of the mean. After polymerization, the nonlinear elastic response was measured using a well-established prestress protocol[47]. Briefly, a constant shear stress $\sigma$ was applied for 30 s, to probe for network creep, and then an oscillatory stress $\delta\sigma$ was superposed with an amplitude of $\sigma/10$ and frequency of 0.5 Hz. The resulting differential strain $\delta\gamma$ was then used to calculate the differential (or tangent) modulus $K' = \delta\sigma/\delta\gamma$. To evaluate network fracture, we applied a linear strain ramp (with a loading rate of 0.5%/s) to the samples after polymerization. The fracture strain of each network was calculated as the strain where the stress reached a peak value. To verify that fracture occurred inside the network rather than at the plates, we performed measurements at different strain rates (0.125 %/s, 1 %/s, 4 %/s) and we compared measurements for bare rheometer plates and for the rheometer plates coated with an adhesive layer by depositing, spreading and drying 30 μL of a solution of fibrinogen (10 mg/mL, Human plasma fibrinogen, Plasminogen, Enzyme Research Laboratories, UK) or poly-L-lysine (Poly-L-lysine solution, 0.1% w/v in H2O, Sigma Aldrich) before depositing the collagen gels[48]. We assessed the degree of plasticity of the collagen networks by performing repeated strain ramps and measuring the hysteresis during each stress-strain cycle. Each strain ramp was performed for strains up to $\gamma_p/2$, where $\gamma_p$ was previously determined for each type of collagen.

**Confocal rheology measurements**

Confocal rheology measurements were performed on a custom-built confocal rheometer consisting of a rheometer head (DSR 301, Anton Paar) mounted in a metal rack on top of an inverted microscope (DMIRB, Leica Microsystems) equipped with a confocal spinning disk (CSU22, Yokogawa Electric Corp.). The collagen networks were imaged using a 488 nm laser (Sapphire 488-30 CHRH, Coherent Inc., Utrecht, Netherlands) for excitation and a back-illuminated cooled EM-CCD camera (C9100, Hamamatsu Photonics, Germany) for detection. As a bottom plate, we used a circular glass coverslip (Menzel Gläser, 40 mm), which was coated beforehand with poly-L-lysine (Sigma Aldrich) to promote attachment of the collagen network to the surface. To determine the gap size, the confocal head was manually lowered towards the glass bottom plate using a micrometer screw until the normal force increased from 0 to 0.02 N, signalling contact with the surface. Subsequently, the rheometer head was raised again from this reference point and the sample was loaded and the gap manually closed to the desired gap size. The experiments were performed at a gap of 0.5 mm and with an upper steel plate with a diameter of 20 mm. Samples were polymerized *in situ* at 22°C and solvent evaporation was prevented by placing a thin layer of mineral oil (Sigma Aldrich) around the measuring geometry. The sample was allowed to polymerize for at least two hours, while the elastic and viscous moduli were monitored by applying small oscillations with a strain amplitude of 0.5% and frequency of 0.5 Hz. Afterwards, a linear strain ramp was applied, analogous to the standard rheology protocol described above, while the network was imaged at a confocal plane located 20 μm above the bottom surface. Time-lapse videos were collected during the strain ramp at a frame rate of 2 fps and exposure time of 500 ms. After fracture (as evident from a drop of the shear stress), we verified that the network had not detached from the lower plate by acquiring a z-stack from the bottom plate upwards, up till a height of 20 μm into the sample, and inspecting the z-stack projection from the side.

**Confocal microscopy imaging**

Confocal data for quantification of the network mesh size were obtained with an inverted Eclipse Ti microscope (Nikon, Tokyo, Japan) using 40× and 100x oil immersion objectives with numerical apertures of 1.30 and 1.49, respectively (Nikon, Tokyo, Japan). The networks were imaged in confocal reflectance mode with a 488 nm Argon laser for illumination (Melles Griot, Albuquerque, NM). Image stacks were acquired starting at a height of 10 μm above the coverslip to avoid surface effects, over a total depth of 20 μm, and with a step size of 0.5 μm. The data are shown in the text as maximum intensity projections obtained with ImageJ[49]. The networks were prepared in dedicated sample holders composed of two coverslips (Menzel™ Microscope Coverslips 24x60mm, #1, Thermo Scientific) separated by a silicon chamber (Grace Bio-Labs CultureWell™ chambered coverglass, Sigma Aldrich). The sample holders were subsequently placed in a petri dish wrapped with humidified tissues and closed by parafilm, in order to prevent sample dehydration, and then placed in a warm room (37°C) or in a temperature-controlled oven (for polymerization at 34°C-30°C-27°C) to allow collagen polymerization for at least two hours before observation.

**Mesh size analysis**

The mesh size of the collagen networks was determined with a custom-written Python code, according to a previously published protocol[50]. Briefly, a z-stack of images was background subtracted, thresholded and binarized with the Otsu method in ImageJ[49] and the Python program was used to count the distance between on and off pixels in each image, for each row and column. The distance distributions were fitted to an exponential function. The characteristic 1/e distance was converted from pixel to micron and taken as the average mesh size.

**Scanning Electron Microscopy**

The samples for scanning electron microscopy were prepared following a previously established protocol[51]. Collagen networks were formed overnight in Eppendorf tubes at 0.5 or 1 mg/mL. The samples were washed three times for 60 minutes each with sodium cacodylate buffer (50 mM cacodylate, 150 mM NaCl, pH 7.4) obtained by mixing Cacodylic acid sodium salt (Sigma Aldrich), 0.2 M HCl and milliQ water. The samples were fixed with 2.5% glutaraldehyde in sodium cacodylate for at least two hours, washed again three times with cacodylate buffer at room temperature, and dehydrated with a stepwise increasing percentage of ethanol in milliQ water by sequential 10-20 min incubations (30, 50, 70, 80, 90, 95 %). Finally, the samples were washed with 50% HDMS (hexamethyldisilazane, Sigma Aldrich) in ethanol and 100% HDMS. HDMS was pipetted out and the Eppendorf tubes were left open overnight in order to dry. Subsequently, the samples were mounted on a support with carbon tape and covered with a 11-14 nm layer of palladium gold with a sputtercoater (Leica EM ACE600). The samples were then imaged with a Scanning Electron Microscope (Verios 460, FEI, Eindhoven, the Netherlands). To determine the fibril diameters, we manually analyzed images taken at a magnification of 80000x, by drawing segments perpendicular to a collagen fibril and then measuring the distance between the edges. To prevent bias, we overlaid a grid of 500x500 nm onto each image through the ImageJ grid plugin, and measured the fibrils at the intersections of the crosses. We note that the measured fibril diameters are semi-quantitative because the sample preparation for electron microscopy introduces fibril shrinkage from drying but also fibril thickening due to metal deposition, while imaging in vacuum also introduces fibril shrinkage. This does not influence our conclusions, because the diameter data are only used as a relative measure between different collagen networks.

**Determination of onset and critical strain for collagen strain-stiffening**

The non-linear rheology data (differential elastic modulus K' as a function of shear stress σ) were evaluated using a custom-written Python routine. The onset stress was determined by considering the minimum value of the experimentally determined K'/σ as a function of σ. The corresponding strain was defined as onset strain $\gamma_o$. To determine the critical strain, we calculated the cubic spline derivative of log K' as a function of log γ. The strain at which this function reached its maximum was defined the critical strain $\gamma_c$. The characteristic strain values are shown as averages with standard error of the mean of at least three independent measurements.

**Determination of structural and mechanical properties of collagen from rheology data**

It was previously shown that the nonlinear elastic response of collagen networks is quantitatively described by a theoretical model of athermal networks of rigid beams[3,31]. Specifically, the increase of the differential modulus *K'* with increasing shear strain γ obeys the following equation of state:

$$\frac{\tilde{\kappa}}{|\Delta\gamma|^{\Phi}} \sim \frac{K'}{|\Delta\gamma|^{f}}\left(\pm 1 + \frac{K'^{\frac{1}{f}}}{|\Delta\gamma|}\right)^{(\Phi - f)}, \qquad (1)$$

where $\tilde{\kappa}$ represents the dimensionless bending rigidity, defined as the ratio between the fiber bending modulus *κ* and stretch modulus μ, $|\Delta\gamma|$ is the distance between the measured strain and the critical strain $\gamma_c$, and ϕ and f are critical exponents controlling the transition from the bend-dominated to the stretch-dominated regime. The critical strain and the critical exponents depend on the network architecture through its average connectivity <*z*> and can be determined from computer simulations of 2D random lattice-based networks[31]. From this same comparison between experiments and simulations, we could also obtain the corresponding values of f and ϕ. Next, we used these parameters as input to fit the experimental strain-stiffening curves to Eq. 1, using the dimensionless bending rigidity of the fibers $\tilde{\kappa}$ as the fitting parameter (See Supplementary Table 2).

**Computer simulations of fiber networks**

We performed athermal and quasistatic shear simulations of elastic networks composed of L by L nodes (or $L_x$ by $L_y$ nodes in cases where the simulation box was asymmetric). In line with previous studies[3,5,31,52], we chose a phantom network architecture that has been proven to capture the essential coarse-grained features controlling the mechanics of collagen networks. The phantomization procedure is as follows: (i) starting from a triangular lattice with unit spacing, only two of the three fibrils meeting at the same node are cross-linked; (ii) for each fibril at least one segment is randomly removed to avoid unphysical system-spanning bonds; (iii) the desired average connectivity $\langle z \rangle$ is achieved by diluting the network, i.e. by randomly removing a fraction 1-$p$ of bonds, yielding $\langle z \rangle = 4p$. In the simplest model considered in this study, each bond was modelled as an elastic linear (Hookean) spring with unit stiffness and rest length equal to the lattice spacing (set to unity). All springs have the same rupture threshold $\lambda$ and break irreversibly when their deformation exceeds $\lambda$. As shown in the Supplementary Information, we also performed simulations where we included a three-body bending potential along straight segments, to test whether the fibril bending stiffness $\kappa$ has any effect on network fracture. These simulations were performed in a similar manner as in previous studies[3], but with the addition of bond rupture events when the axial deformation of a spring exceeded $\lambda$. To test for the influence of fibril plasticity, we finally performed simulations where we allowed for spring lengthening: as soon as the spring length reaches $l_m$, its rest length increases to $r_2$. The relation $l_m + r_2 - 2r_1 < \lambda$ must hold, otherwise spring lengthening would occur after bond rupture (inaccessible region above the line in Fig. 4d). In all cases, networks were subjected to a simple shear deformation. After each small increment of the shear strain $\gamma$, the energy was minimized using the FIRE algorithm[53], with a tolerance (in reduced units) $F_{rms} \leq 10^{-5}$, ensuring that simulations were carried out in the athermal limit. Bonds were broken one at a time (if any), starting from the weakest (the one that exceeded $\lambda$ the most) and performing energy minimization in between fracture events. Lee-Edwards boundary conditions were employed, except in the case of strongly asymmetric boxes ($L_x >> L_y$), for which the top and bottom boundary nodes were moved rigidly (to avoid unphysical elastic waves travelling through the system when bonds are fracturing). Laterally (in the x-direction), standard periodic boundary conditions were always employed. The mechanical response was quantified via the shear stress $\sigma$ calculated using the $xy$-component of the virial stress tensor. Quantities ($\sigma$ and $K'$) were expressed in reduced units and averaged over a sufficiently large number of configurations, ranging from 500 for L=24 to 20 for L=256.

## REFERENCES

1. Mouw, J. K., Ou, G. & Weaver, V. M. Extracellular matrix assembly: a multiscale deconstruction. *Nat. Rev. Mol. Cell Biol.* **15**, 771–785 (2014).

2. Maxwell, J. C. On the calculation of the equilibrium and stiffness of frames. *London, Edinburgh, Dublin Philos. Mag. J. Sci.* **27**, 294–299 (1864).

3. Sharma, A. *et al.* Strain-controlled criticality governs the nonlinear mechanics of fibre networks. *Nat. Phys.* **12**, 584–587 (2016).

4. Burla, F., Mulla, Y., Vos, B. E., Aufderhorst-Roberts, A. & Koenderink, G. H. From mechanical resilience to active material properties in biopolymer networks. *Nat. Rev. Phys.* **1**, 249–263 (2019).

5. Licup, A. J. *et al.* Stress controls the mechanics of collagen networks. *Proc. Natl. Acad. Sci.* **112**, 9573–9578 (2015).

6. LaCroix, A. S., Duenwald-Kuehl, S. E., Lakes, R. S. & Vanderby, R. Relationship between tendon stiffness and failure: A metaanalysis. *J. Appl. Physiol.* **115**, 43–51 (2013).

7. Converse, M. I. *et al.* Detection and characterization of molecular-level collagen damage in overstretched cerebral arteries. *Acta Biomater.* **67**, 307–318 (2018).

8. Anne M. Robertson, Xinjie Duan, Khaled M. Aziz, Michael R. Hill, Simon C. Watkins, and J. R. C. Diversity in the Strength and Structure of Unruptured Cerebral Aneurysms. *Ann Biomed Eng.* **25**, 289–313 (2016).

9. Sugita, S. & Matsumoto, T. Local distribution of collagen fibers determines crack initiation site and its propagation direction during aortic rupture. *Biomech. Model. Mechanobiol.* **17**, 577–587

(2018).

10. Tran, Q. D., Marcos & Gonzalez-Rodriguez, D. Permeability and viscoelastic fracture of a model tumor under interstitial flow. *Soft Matter* **14**, 6386–6392 (2018).

11. Arroyo, M. & Trepat, X. Hydraulic fracturing in cells and tissues: fracking meets cell biology. *Curr. Opin. Cell Biol.* **44**, 1–6 (2017).

12. Veres, S. P., Harrison, J. M. & Lee, J. M. Repeated subrupture overload causes progression of nanoscaled discrete plasticity damage in tendon collagen fibrils. *J. Orthop. Res.* **31**, 731–737 (2013).

13. Herod, T. W., Chambers, N. C. & Veres, S. P. Collagen fibrils in functionally distinct tendons have differing structural responses to tendon rupture and fatigue loading Acta Biomaterialia Collagen fibrils in functionally distinct tendons have differing structural responses to tendon rupture and fati. *Acta Biomater.* **42**, 296–307 (2016).

14. Fratzl, P. & Weinkamer, R. Nature's hierarchical materials. *Prog. Mater. Sci.* **52**, 1263–1334 (2007).

15. Buehler, M. J. & Yung, Y. C. Deformation and failure of protein materials in physiologically extreme conditions and disease. *Nat. Mater.* **8**, 175–188 (2009).

16. Svensson, R. B., Mulder, H., Kovanen, V. & Magnusson, S. P. Fracture mechanics of collagen fibrils: Influence of natural cross-links. *Biophys. J.* **104**, 2476–2484 (2013).

17. Svensson, R. B., Smith, S. T., Moyer, P. J. & Magnusson, S. P. Effects of maturation and advanced glycation on tensile mechanics of collagen fibrils from rat tail and Achilles tendons. *Acta Biomater.* **70**, 270–280 (2018).

18. Misof, K., Landis, W. J., Klaushofer, K. & Fratzl, P. Collagen from the osteogenesis imperfecta mouse model (oim) shows reduced resistance against tensile stress. *J. Clin. Invest.* **100**, 40–45 (1997).

19. Ribeiro, J. F., dos Anjos, E. H. M., Mello, M. L. S. & de Campos Vidal, B. Skin Collagen Fiber Molecular Order: A Pattern of Distributional Fiber Orientation as Assessed by Optical Anisotropy and Image Analysis. *PLoS One* **8**, 5–7 (2013).

20. Meng, Q. *et al.* Journal of the Mechanical Behavior of Biomedical Materials The effect of collagen fi bril orientation on the biphasic mechanics of articular cartilage. *J. Mech. Behav. Biomed. Mater.* **65**, 439–453 (2017).

21. Bos, K. J. *et al.* Collagen fibril organisation in mammalian vitreous by freeze etch/rotary shadowing electron microscopy. *Micron* **32**, 301–306 (2001).

22. Lindeman, J. H. N. *et al.* Distinct defects in collagen microarchitecture underlie vessel-wall failure in advanced abdominal aneurysms and aneurysms in Marfan syndrome. *Proc. Natl. Acad. Sci. U. S. A.* **107**, 862–865 (2010).

23. Vanderheiden, S. M., Hadi, M. F. & Barocas, V. H. Crack Propagation Versus Fiber Alignment in Collagen Gels: Experiments and Multiscale Simulation. *J. Biomech. Eng.* **137**, (2015).

24. Hadi, M. F. & Barocas, V. H. Microscale Fiber Network Alignment Affects Macroscale Failure Behavior in Simulated Collagen Tissue Analogs. *J. Biomech. Eng.* **135**, 021026 (2013).

25. Dhume, R. Y., Shih, E. D. & Barocas, V. H. Multiscale model of fatigue of collagen gels. *Biomech. Model. Mechanobiol.* **18**, 175–187 (2019).

26. Bircher, K., Zündel, M., Pensalfini, M., Ehret, A. E. & Mazza, E. Tear resistance of soft collagenous tissues. *Nat. Commun.* **10**, 1–13 (2019).

27. Leikin, S., Rau, D. C. & Parsegian, V. A. Temperature-favoured assembly of collagen is driven by hydrophilic not hydrophobic interactions. *Nat. Struct. Biol.* **2**, 205–210 (1995).

28. Traub, W. Molecular assembly in collagen. *FEBS Lett.* **92**, 114–120 (1978).

29. Guthold, M. *et al.* A comparison of the mechanical and structural properties of fibrin fibers with other protein fibers. *Cell Biochem. Biophys.* **49**, 165–181 (2007).

30. Asif Iqbal, S. M., Deska-gauthier, D. & Kreplak, L. Assessing collagen fibrils molecular damage after a single stretch–release cycle. *Soft Matter* (2019). doi:10.1039/c9sm00832b

31. Jansen, K. A. *et al.* The Role of Network Architecture in Collagen Mechanics. *Biophys. J.* **114**, 2665–2678 (2018).

32. Shayegan, M., Altindal, T., Kiefl, E. & Forde, N. R. Intact Telopeptides Enhance Interactions between Collagens. *Biophysj* **111**, 2404–2416 (2016).

33. Zhang, L., Rocklin, D. Z., Sander, L. M. & Mao, X. Fiber networks below the isostatic point: Fracture without stress concentration. *Phys. Rev. Mater.* **1**, 1–5 (2017).

34. Driscoll, M. M. *et al.* The role of rigidity in controlling material failure. *Proc. Natl. Acad. Sci. U. S. A.* **113**, 10813–10817 (2016).

35. Berthier, E. *et al.* Rigidity percolation control of the brittle-ductile transition in disordered networks. *Phys. Rev. Mater.* **075602**, 1–9 (2019).

36. Dussi, S., Tauber, J. & van der Gucht, J. Athermal Fracture of Elastic Networks: How Rigidity Challenges the Unavoidable Size-Induced Brittleness. 1–13 (2019). arXiv:1907.11466

37. Arevalo, R. C., Urbach, J. S. & Blair, D. L. Size-dependent rheology of type-I collagen networks. *Biophys. J.* **99**, 8–11 (2010).

38. Munster, S. *et al.* Strain history dependence of the nonlinear stress response of fibrin and collagen networks. *Proc. Natl. Acad. Sci.* **110**, 12197–12202 (2013).

39. Ban, E. *et al.* Mechanisms of Plastic Deformation in Collagen Networks Induced by Cellular Forces. *Biophys. J.* **114**, 450–461 (2018).

40. Liu, J. *et al.* Energy dissipation in mammalian collagen fibrils: Cyclic strain-induced damping, toughening, and strengthening. *Acta Biomater.* **80**, 217–227 (2018).

41. Depalle, B., Qin, Z., Shefelbine, S. J. & Buehler, M. J. Influence of cross-link structure, density and mechanical properties in the mesoscale deformation mechanisms of collagen fibrils. *J. Mech. Behav. Biomed. Mater.* **52**, 1–13 (2015).

42. Fratzl, P. *et al.* Fibrillar structure and mechanical properties of collagen. *J. Struct. Biol.* **122**, 119–122 (1998).

43. Puxkandl, R. *et al.* Viscoelastic properties of collagen: Synchrotron radiation investigations and structural model. *Philos. Trans. R. Soc. B Biol. Sci.* **357**, 191–197 (2002).

44. Prince, E. & Kumacheva, E. Design and applications of man-made biomimetic fibrillar hydrogels. *Nat. Rev. Mater.* **4**, 99–115 (2019).

45. Ricard-Blum, S. The Collagen Family. *Cold Spring Harb. Perspect. Biol.* **3**, 1–19 (2011).

46. Aper, S. J. A. *et al*. Colorful protein-based fluorescent probes for collagen imaging. *PLoS One* **9**, 1–21 (2014).

47. Broedersz, C. P. *et al*. Measurement of nonlinear rheology of cross-linked biopolymer gels. *Soft Matter* **6**, 4120–4127 (2010).

48. Nam, S., Hu, K. H., Butte, M. J. & Chaudhuri, O. Strain-enhanced stress relaxation impacts nonlinear elasticity in collagen gels. *Proc. Natl. Acad. Sci.* **113**, 5492–5497 (2016).

49. Schindelin, J. *et al*. Fiji: an open-source platform for biological-image analysis. *Nat. Methods* **9**, 676 (2012).

50. Kaufman, L. J. *et al*. Glioma Expansion in Collagen I Matrices : Analyzing Collagen Concentration-Dependent Growth and Motility Patterns. *Biophys. J.* **89**, 635–650 (2005).

51. Baradet, T. C., Haselgrove, J. C. & Weisel, J. W. Three-dimensional reconstruction of fibrin clot networks from stereoscopic intermediate voltage electron microscope images and analysis of branching. *Biophys. J.* **68**, 1551–1560 (1995).

52. Burla, F., Tauber, J., Dussi, S., van der Gucht, J. & Koenderink, G. H. Stress management in composite biopolymer networks. *Nature Physics* **15**, (2019).

53. Bitzek, E., Koskinen, P., Gähler, F., Moseler, M. & Gumbsch, P. Structural relaxation made simple. *Phys. Rev. Lett.* **97**, (2006).

## Figure captions

**Figure 1: Fracture of reconstituted collagen networks under shear deformation.** (a) The experimental setup consists of a parallel plate rheometer with a steel top plate to apply a shear strain $\gamma$ and a transparent glass bottom plate, mounted on an inverted spinning disc confocal microscope. We image a horizontal (xy) plane at a fixed distance (20 μm) away from the bottom surface (yellow-shaded region) and shifted by half of the plate radius from the centre of the sample. Thus, the local strain is 50% of the strain at the edge, which is reported by the rheometer. (b) Imposed linear strain ramp (top) and example measurement of the resulting shear stress $\sigma$ (blue, left-hand y-axis) as a function of shear strain $\gamma$ for a 1 mg/mL network of bovine dermal telocollagen polymerized at 25°C. The total fluorescence intensity of the confocal plane is also shown (pink, right-hand y-axis). (c) Confocal fluorescence images of the network labelled with eGFP-CNA35, at various strain levels (see legend) that correspond to the circles in panel (b). Scale bar is 10 μm. The arrow labelled 'shear' indicates the direction of shear. The data represent one repeat; more data are shown in the Supplementary Information.

**Figure 2: The strain at which collagen networks fracture varies with network structure.** (a) Top rows: confocal reflectance images of collagen networks (scale bar: 10 μm); bottom rows: corresponding scanning electron microscopy images (scale bar: 1 μm). The networks were assembled from collagens from different animals, tissues, and at different temperatures (see Supplementary Table 1). (b) Collagen networks possess a hierarchical structure that can differ at the *network* level (mesh size), *fibril-fibril interaction* level (junctions formed either by branching or fibril-fibril crossings), and *fibril* level (diameter) and *molecular* level (intrafibrillar crosslinking via telopeptide end regions). (c) Overview of peak strains and stresses at fracture for the entire range of collagen networks. Same symbol shapes indicate same animal and tissues, open symbols refer to collagens without telopeptides (uncrosslinked fibrils) while closed symbols refer to telocollagens (crosslinked fibrils). Peak stresses and strains are shown as averages of at least three independent repeats ± S.E.M.

**Figure 3: Fracture experiments on collagen networks and simulations on subisostatic spring networks both show that the average network connectivity governs the rupture strain.** (a) Experimentally measured (same symbols as in Figure 2c) and simulated (lines) peak strain as a function of the average connectivity <*z*>. The simulations were performed for two fibril rupture thresholds ($\lambda$=10%, lower curve, and $\lambda$=20%, upper curve) that bracket the physiologically relevant range. The inset shows the correlation between the measured critical strain $\gamma_c$ and the peak strain $\gamma_p$. Experimentally, <*z*> is inferred from $\gamma_c$ by mapping the strain-stiffening response on simulations. (b) Example simulations of network fracture. Upper panels: simulation snapshots for a network with <*z*>=3.2 and $\lambda$=10% at shear strains of 20% (before fracture), 30% (peak stress) and 40% (post-rupture). Springs are colored according to their axial deformation (grey to pink, low to high). Lower panel: stress-strain curves for large networks (L=256) with <*z*>=3.7 and $\lambda$=10% (pink), <*z*>=3.2 and $\lambda$=10% (solid blue), and <*z*>=3.2 and $\lambda$=20% (dashed blue). Black circles correspond to the snapshots in the upper panel.

**Figure 4: The influence of structural plasticity on the fracture of collagen networks.** (a) Plastic effects are implemented in the model by allowing spring lengthening based on two control parameters: the onset axial strain for fibril lengthening $l_m$ and the increased rest length $r_2$. Experimentally, plasticity can occur at the network-level due to branch/bundle opening (top schematic) and at the fiber-level due to fiber lengthening (bottom schematic). (b) Experimental shear rheology cycles measured for three different collagen networks (see legend). The normalized dissipated work *W* calculated from the areas under the loading and unloading curves according to $W = (A_{load}-A_{unload})/(A_{load}+A_{unload})$ is indicated next to each curve. Larger values indicate larger dissipation, meaning larger plasticity. (c) Direct comparison of normalized stress-strain curves from experiments (solid line, averaged over 3 samples) and from simulations (averaged over 20 configurations with *L*=256). Top: the response of telocollagen (Telo bovine) networks is captured by simulations of fibers that do not lengthen. Middle: the response of atelocollagen (Atelo bovine) networks reveals significant lengthening. Bottom: for telocollagen networks with low connectivity (Rat 37), the experimental curve matches with simulations only if we assume either an unrealistically large fiber rupture threshold $\lambda$ or fiber lengthening. (d) Network peak strain $\gamma_p$ of simulated networks (*L*=256, *z*=2.9, $\lambda$=30%, representative of Rat37) as a function of $l_m$ and $r_2$. Symbols are color-coded according to the value of $\gamma_p$ (see color bar). Up triangles indicate networks with an enhanced $\gamma_p$ compared to the elastic limit (no lengthening) while down triangles indicate networks with a reduced $\gamma_p$. Blue line is a guide-to-the-eye separating the two regions. The shaded grey area is inaccessible, because bond lengthening would occur after fibre rupture.

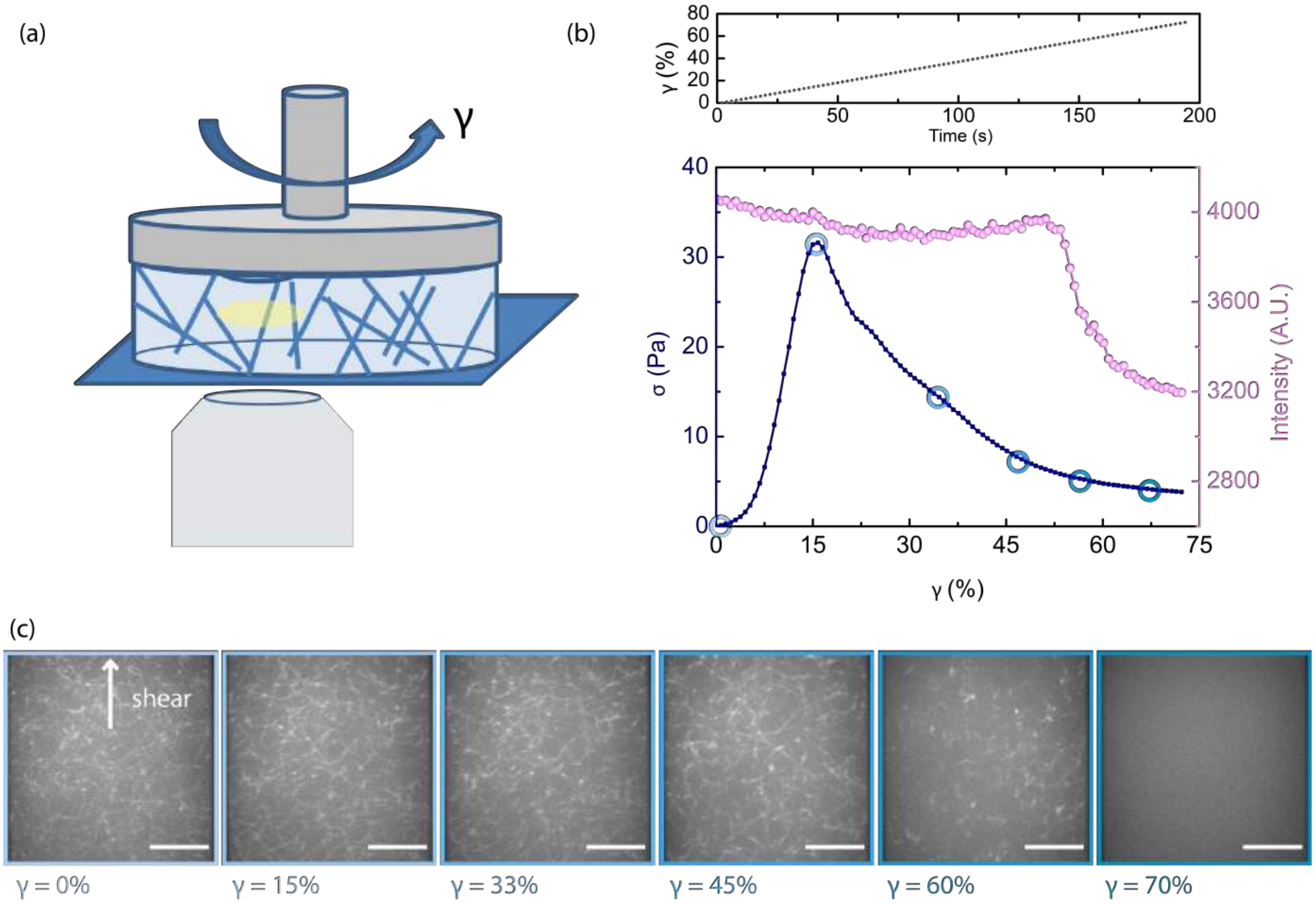

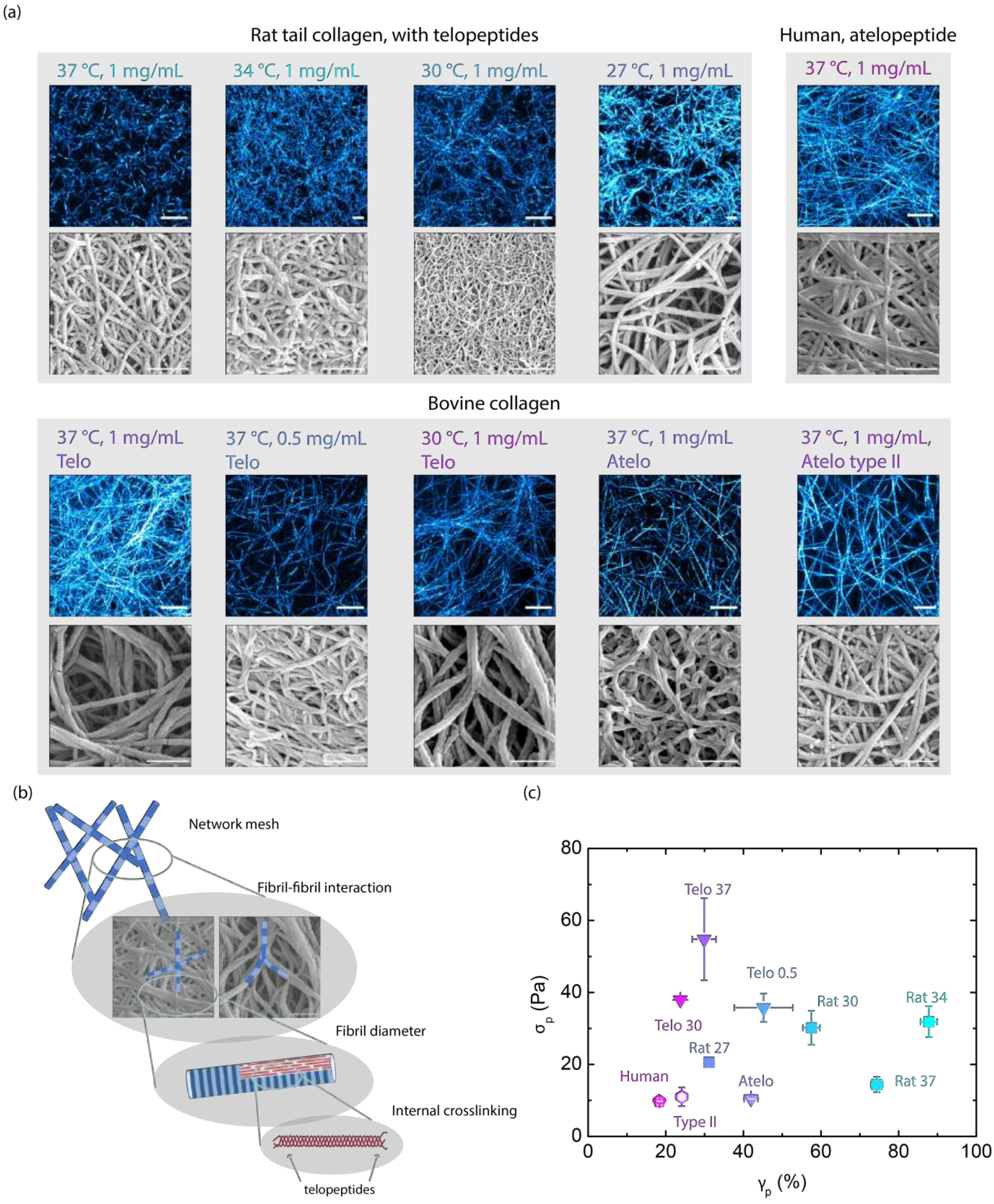
(a)
Rat tail collagen, with telopeptides
Human, atelopeptide
37 °C, 1 mg/mL
34 °C, 1 mg/mL
30 °C, 1 mg/mL
27 °C, 1 mg/mL
37 °C, 1 mg/mL
Bovine collagen
37 °C, 1 mg/mL Telo
37 °C, 0.5 mg/mL Telo
30 °C, 1 mg/mL Telo
37 °C, 1 mg/mL Atelo
37 °C, 1 mg/mL, Atelo type II
(b)
Network mesh
Fibril-fibril interaction
Fibril diameter
Internal crosslinking
telopeptides
(c)
Telo 37
Telo 0.5
Telo 30
Rat 30
Rat 34
Rat 27
Human
Atelo
Rat 37
Type II
$\sigma_p$ (Pa)
$\gamma_p$ (%)
0
20
40
60
80
100

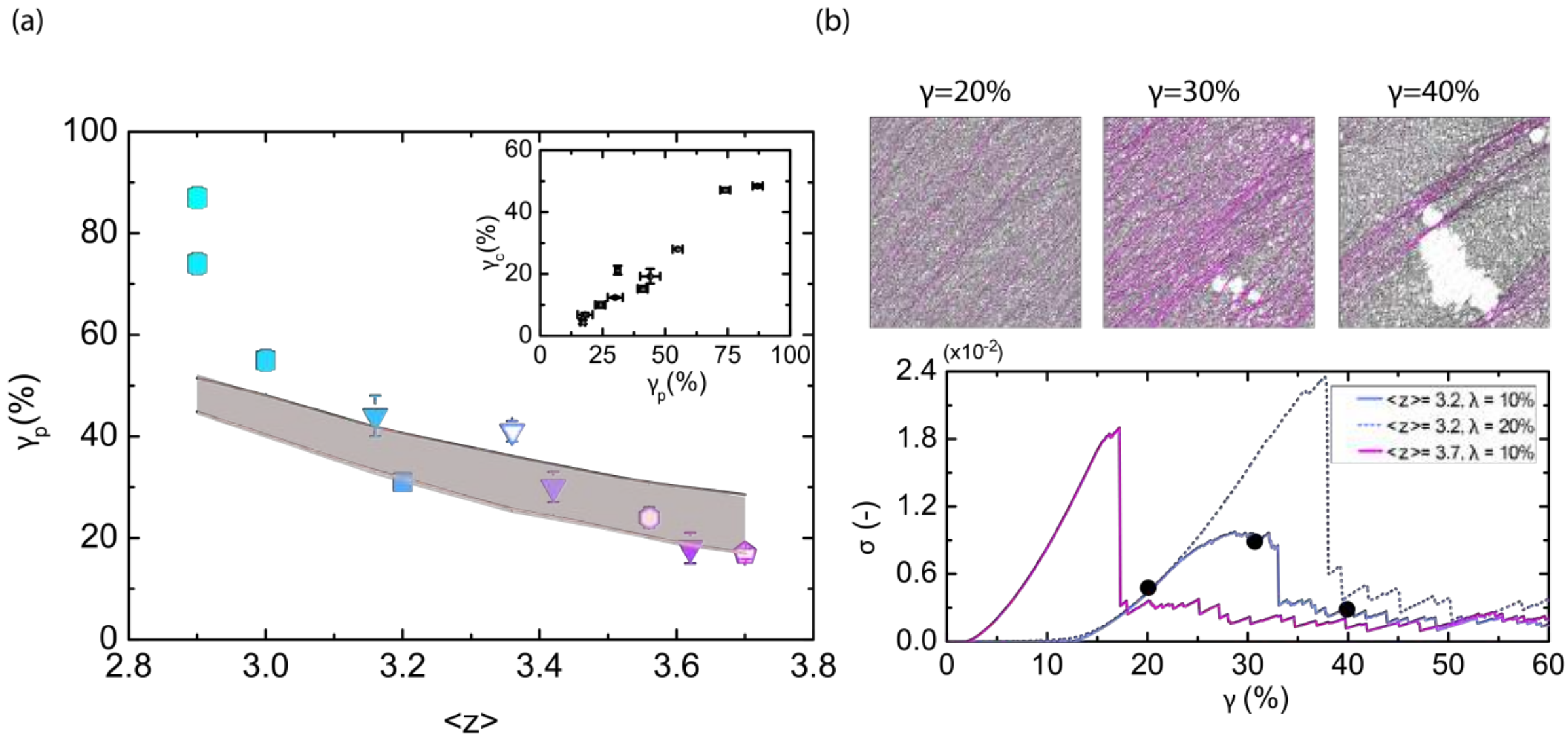

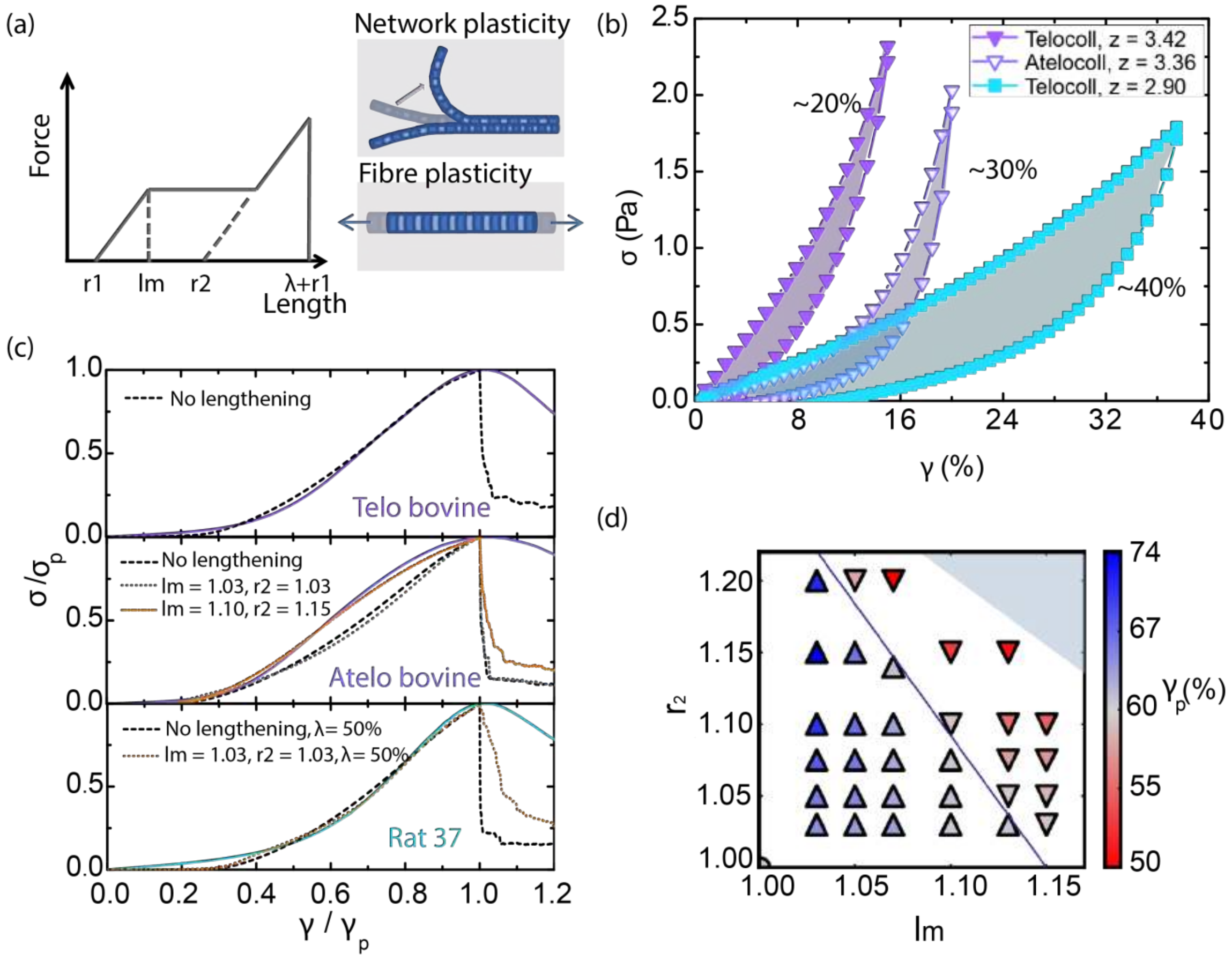
(a)
Force
r1
lm
r2
λ+r1
Length
Network plasticity
Fibre plasticity
(b)
Telocoll, z = 3.42
Atelocoll, z = 3.36
Telocoll, z = 2.90
~20%
~30%
~40%
σ (Pa)
γ (%)
(c)
No lengthening
Telo bovine
No lengthening
lm = 1.03, r2 = 1.03
lm = 1.10, r2 = 1.15
Atelo bovine
No lengthening, λ= 50%
lm = 1.03, r2 = 1.03, λ= 50%
Rat 37
σ /σp
γ / γp
(d)
r2
lm
γp(%)

# Supplementary Information

## Disorder protects collagen networks from fracture

F. Burla[1]†, S. Dussi[2]†, C. Martinez-Torres[1,3], J. Tauber[2], J. van der Gucht[2]*, G.H. Koenderink[1,3]*

[1]*AMOLF, Department of Living Matter, Biological Soft Matter group, Science Park 104, 1098 XG Amsterdam, the Netherlands*

[2]*Physical Chemistry and Soft Matter, Wageningen University & Research, Stippeneng 4, 6708 WE Wageningen, Netherlands.*

[3] *Department of Bionanoscience, Kavli Institute of Nanoscience Delft, Delft University of Technology, Van der Maasweg 9, 2629 HZ Delft, the Netherlands*

*† These authors contributed equally to this work.*

** Corresponding authors: g.h.koenderink@tudelft.nl, jasper.vandergucht@wur.nl*

**List of Supplementary Tables**

**List of Supplementary Figures**

18. Atelocollagen (uncrosslinked) networks are softer than telocollagen (crosslinked) networks and exhibit a larger softening interval (experiments).
19. Dependence of the shape of the stress-strain response in simulations on the simulation parameters.
20. The shape of the stress-strain response of collagen networks in simulations and experiments reveals differences in plasticity depending on network connectivity and intrafibrillar crosslinking.
21. Plasticity in collagen networks inferred from matching experiments with simulations performed with different fiber rupture and lengthening parameters.

**Table S1: Overview of the collagen sources used in the experiments.** Specification of the animal source, tissue origin, collagen type (either type I or type II), presence of telopeptide end regions that mediate intrafibrillar crosslinking, stock concentration, stock buffer conditions, and the supplier.

| Name | Animal source | Tissue origin | Collagen type | Telopeptides (yes/no) | Stock conc. (mg/ml) | Buffer | Supplier |
|---|---|---|---|---|---|---|---|
| PureCol | Bovine | Dermis | I | No | 3.2 | 0.01 M HCl | CellSystems |
| TeloCol | Bovine | Dermis | I | Yes | 3 | 0.01 M 0.02 HCl | CellSystems |
| RatCol | Rat | Tail tendons | I | Yes | 8.34 | 0.02 M acetic acid | Corning |
| VitriCol | Human | Neonatal cells | I | No | 3 | 0.01 M HCl | CellSystems |
| Type II | Bovine | Tracheal cartilage | II | No | 3 | 0.02 M acetic acid | Sigma Aldrich |

**Table S2: Overview of parameters quantifying the strain-stiffening response of the entire range of collagen networks.** Collagen networks were assembled from different collagen types (I or II) and tissue and animal sources (Supplementary Table S1), at different temperatures (27, 30, 34, or 37$^o$C) and at different concentrations (1 mg/ml if unspecified, otherwise 0.5 mg/ml). First, the average connectivity $<z>$ was determined by calibrating the critical strain $\gamma_c$ against 2D network simulations [1]. Next, the corresponding critical exponents f and Φ were determined from the best-matching simulations. Finally, the dimensionless bending rigidity $\tilde{\kappa}$ was determined by fitting the constitutive equation for fibrous networks (Eq. 1 in the Materials and Methods section of the main text) to the experimentally measured strain-stiffening curves of collagen networks (see Supplementary Figure S8a) using $\tilde{\kappa}$ as the only adjustable parameter. Data are averages over 3 independent measurements.

| **Collagen types** | **$\gamma_c$ [1]** | **$<z>$** | **F** | **Φ** | **$\tilde{\kappa}$ *$10^4$ (average±s.e.m.)** |
|---|---|---|---|---|---|
| Telo 0.5 (Bovine telo 37$^o$C 0.5 mg/mL) | 0.19 ± 0.02 | 3.16 | 0.74 | 2.1 | 2.25 ± 0.85 |
| Human (Human atelo 37$^o$C 1.0 mg/mL) | 0.045 ± 0.006 | 3.7 | 0.83 | 2 | 0.4 ± 0.08 |
| Rat 37 (Rat tail telo 37$^o$C 1.0 mg/mL) | 0.47 ± 0.007 | 2.9 | 0.77 | 2.2 | 1.7± 0.15 |
| Type II (Bovine atelo typeII 37$^o$C 1.0 mg/mL) | 0.1 ± 0.001 | 3.56 | 0.81 | 2 | 0.33 ± 0.07 |
| Rat 27 (Rat tail telo 27$^o$C 1.0 mg/mL) | 0.21 ± 0.01 | 3.2 | 0.75 | 2.1 | 0.95 ± 0.05 |
| Rat 34 (Rat tail telo 34$^o$C 1.0 mg/mL) | 0.48 ± 0.005 | 2.9 | 0.77 | 2.2 | 1.6 ± 0.15 |
| Rat 30 (Rat tail telo 30$^o$C 1.0 mg/mL) | 0.28 ± 0.01 | 3 | 0.7 | 2.2 | 1.05 ± 0.06 |
| Telo 37 (Bovine telo 37$^o$C 1.0 mg/mL) | 0.12 ± 0.003 | 3.42 | 0.77 | 2.1 | 1.4 ± 0.1 |
| Atelo (Bovine atelo 37$^o$C 1.0 mg/mL) | 0.15 ± 0.01 | 3.36 | 0.77 | 2.1 | 0.37± 0.02 |
| Telo 30 (Bovine telo 30$^o$C 1.0 mg/mL) | 0.068 ± 0.006 | 3.62 | 0.82 | 2 | ~1* |

*difficult to have a good fit with this collagen type

(a)

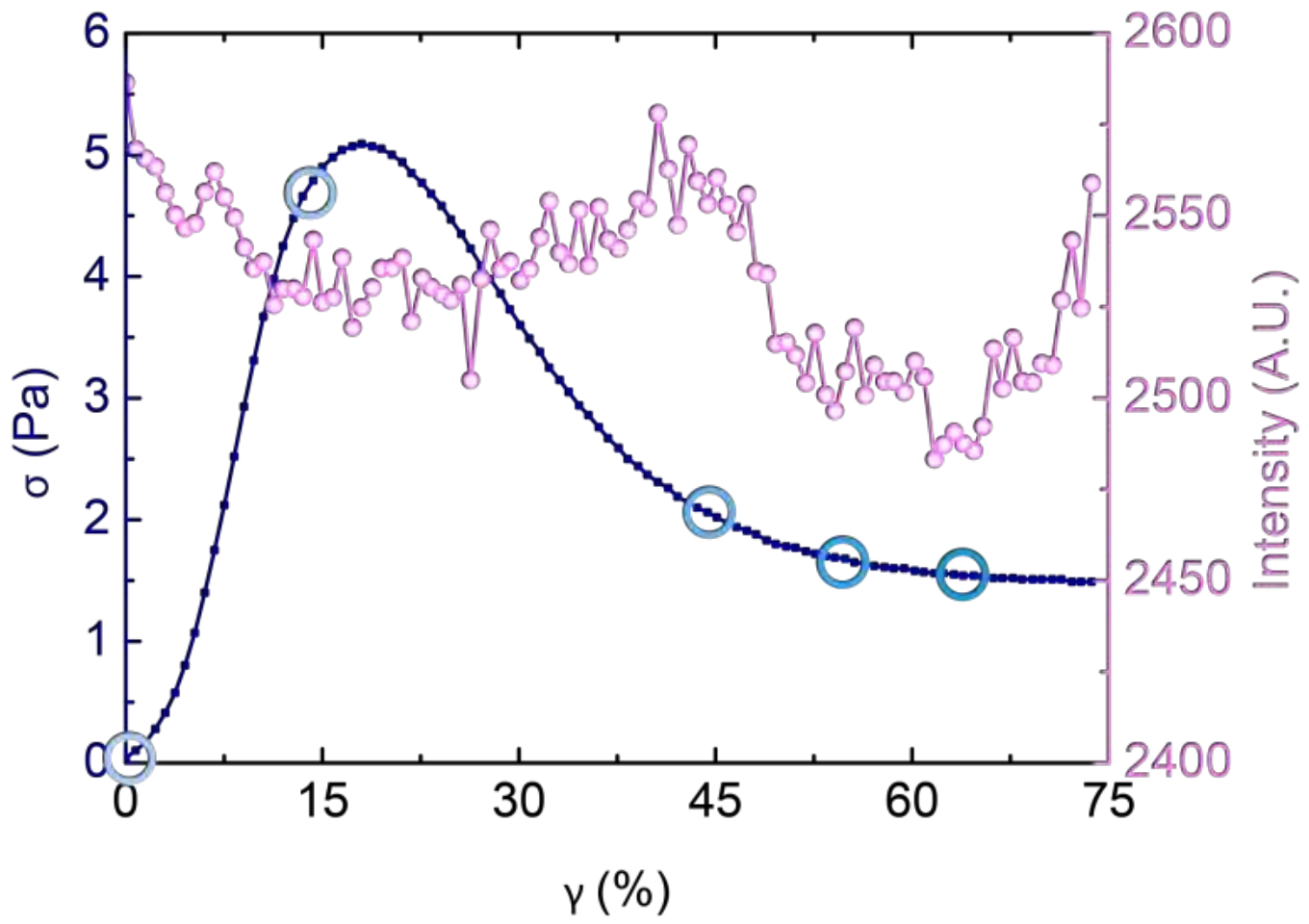

(b)

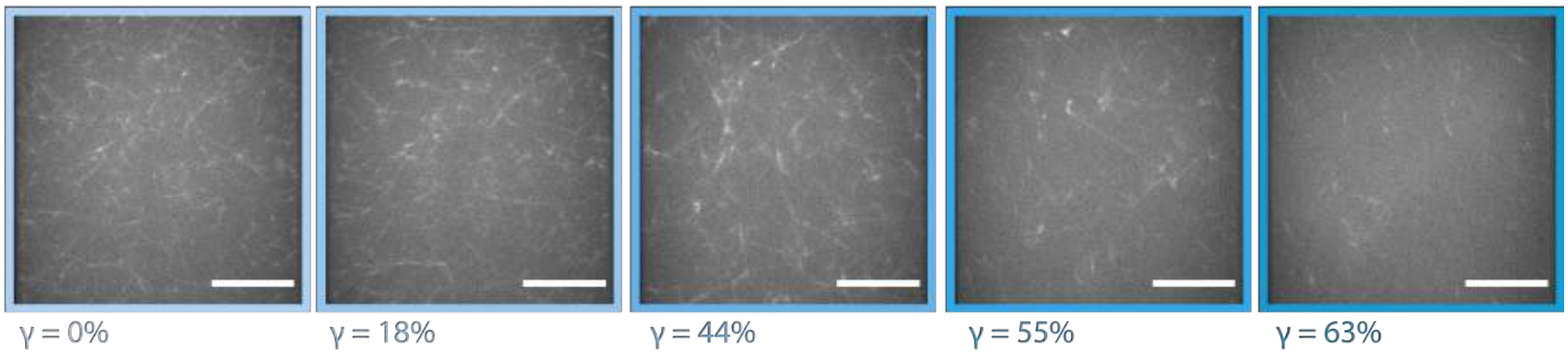

**Supplementary Figure S1: Shear-induced fracture of a bovine telocollagen network visualized with confocal rheology.** Analogously to Figure 1 in the main text, panel (a) shows the stress-strain curve recorded with the rheometer (blue, left-hand y-axis) and the total intensity of the confocal plane imaged by spinning disc confocal microscopy (pink, right-hand y-axis), while panel (b) shows selected snapshots of the network (single confocal slices) at different levels of applied shear strain (see legend). The measurement was performed for a 1 mg/mL bovine dermal telocollagen network polymerized at 22°C and fluorescently labelled with CNA35. Scale bar is 10 µm.

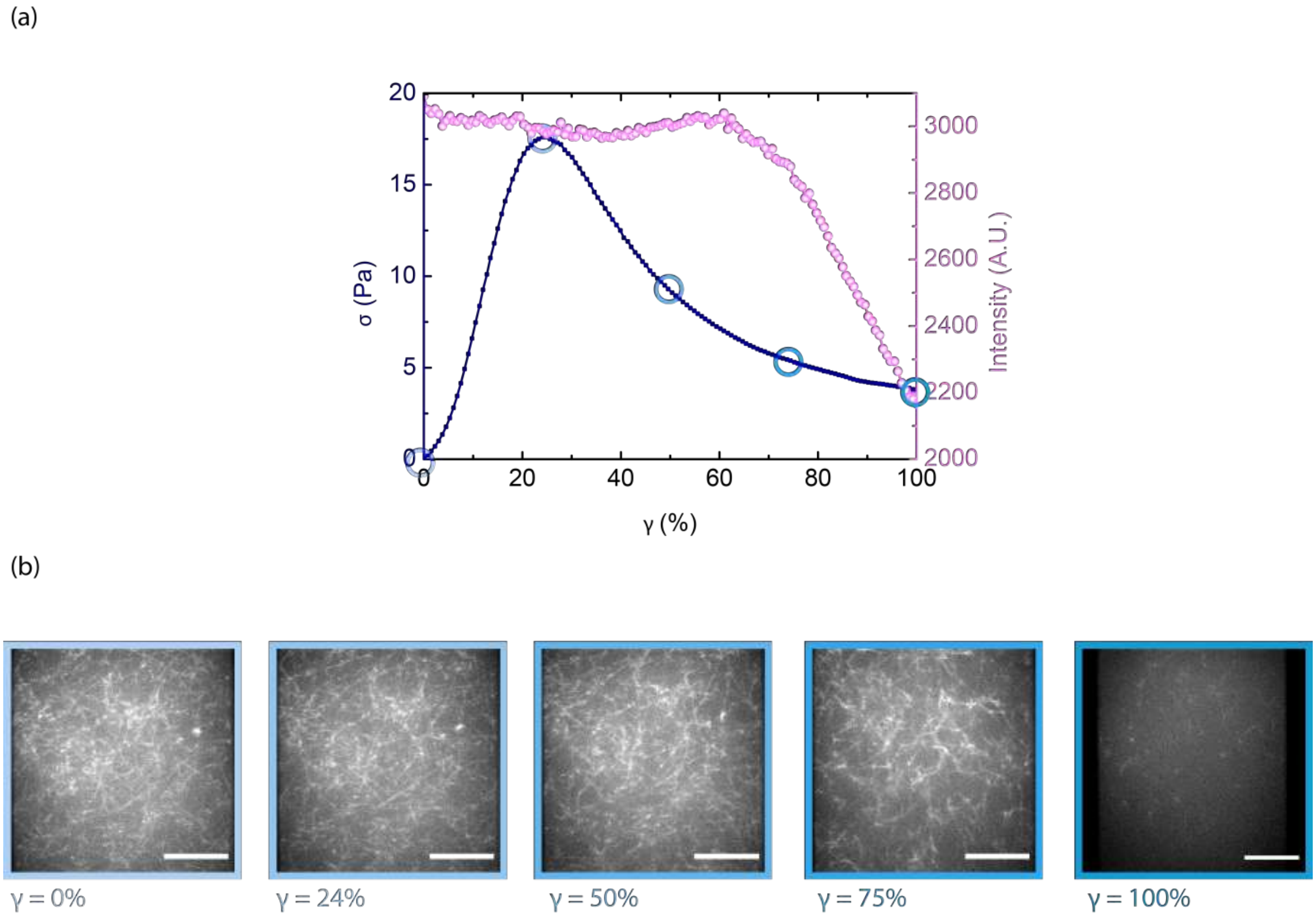

**Supplementary Figure S2: Shear-induced fracture of a rat tail collagen network visualized with confocal rheology.** (a) Stress-strain curve recorded with the rheometer (blue, left-hand y-axis) and total fluorescence intensity of a confocal slice recorded by spinning disc confocal microscopy (pink, right-hand y-axis). (b) Selected snapshots (single confocal slices) of the network obtained at a distance of 20 µm above the bottom surface at different levels of applied shear strain (see legend). The measurement was performed for a 1 mg/mL rat tail telocollagen network polymerized at 22°C and fluorescently labelled with CNA35. Scale bar is 10 µm.

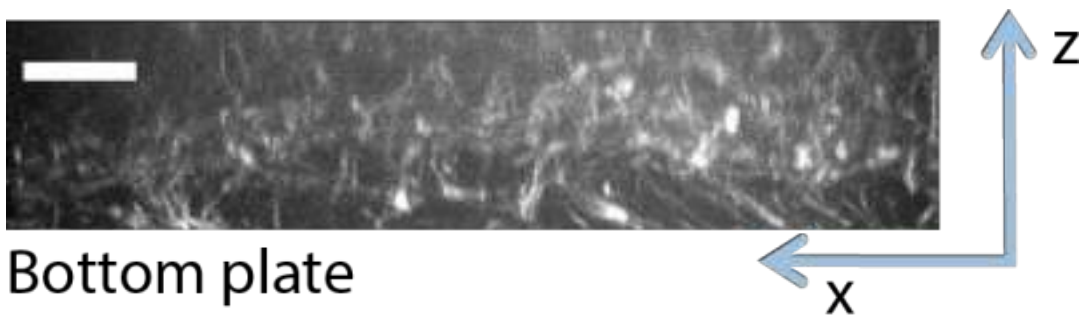

**Supplementary Figure S3: Post-fracture confocal fluorescence z-stack of a collagen network.** Y-projection of a confocal z-stack recorded by spinning disc confocal microscopy for a bovine telopeptide collagen network after shear-induced fracture. The scale bar is 5 µm in the x direction and the total z height is 20 µm. The presence of connected fibers on the bottom plate suggests that network failure is cohesive and not due to detachment from the lower surface.

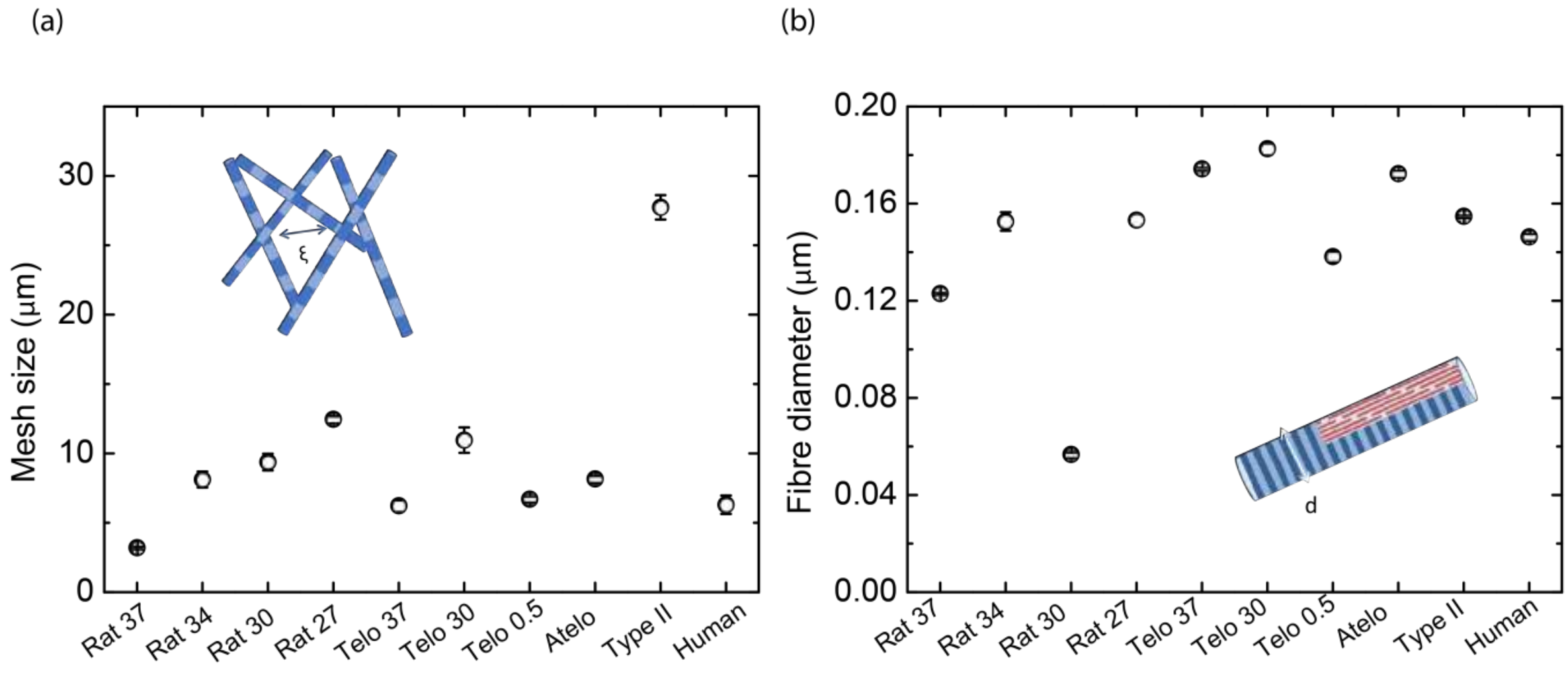

**Supplementary Figure S4: Mesh size and fiber diameters of the collagen networks.** (a) Mesh size ξ for the entire range of collagen networks obtained from confocal reflectance images. Data are averages over measurements on three independently prepared samples, and error bars represent the standard error of the mean. (b) Corresponding average fiber diameter *d* obtained from scanning electron microscopy images. Data are averages over at least 100 different fibers in different regions of each sample (one sample per condition), and error bars represent standard error of the mean.

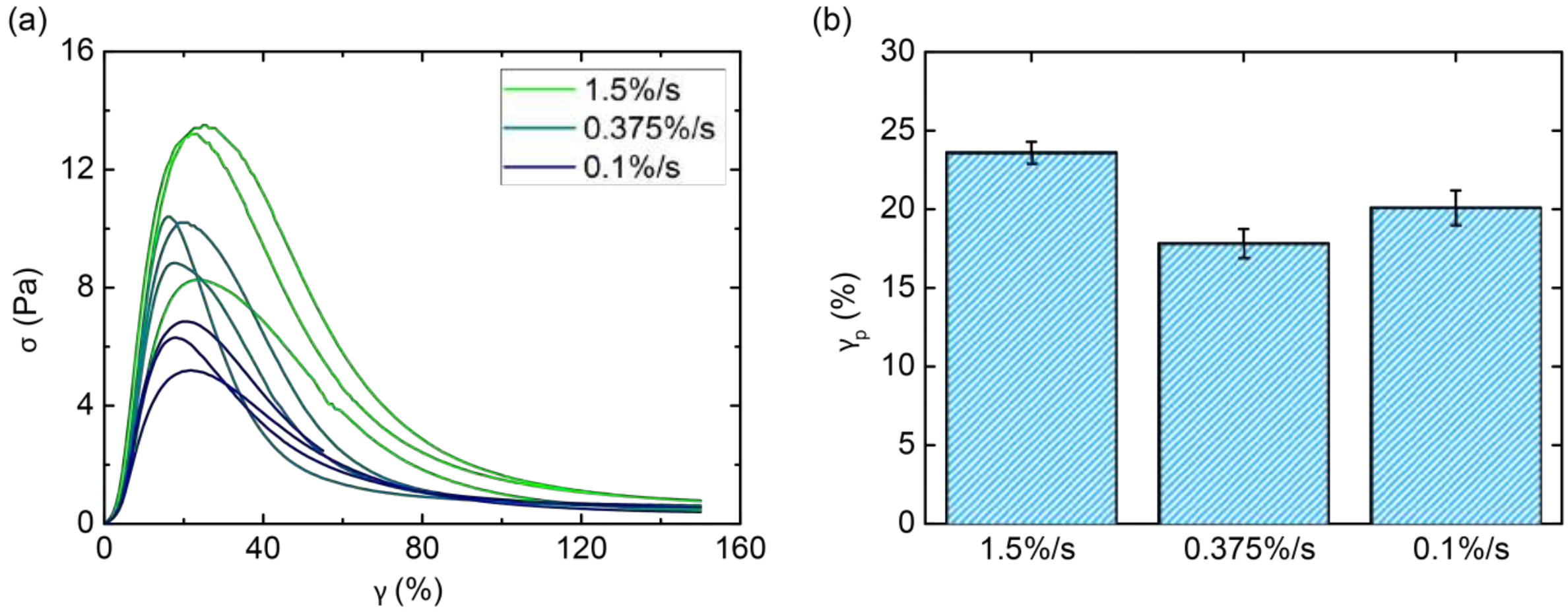

**Supplementary Figure S5: The peak strain of collagen networks is independent of shear rate.** (a) Stress-strain curves measured for 1 mg/mL human atelocollagen networks during strain ramps performed at different strain rates between 0.1%/s and 1.5 %/s (see legend). The peak strain $\gamma_p$ shows little variation, while the peak stress decreases with decreasing strain rate, possibly due to plastic effects. Each curve represents a single experiment. (b) Peak strain $\gamma_p$ for the three different strain rates, showing averages of three independent measurements ± S.E.M. Note that 0.375%/s is the strain rate used for the majority of our experiments.

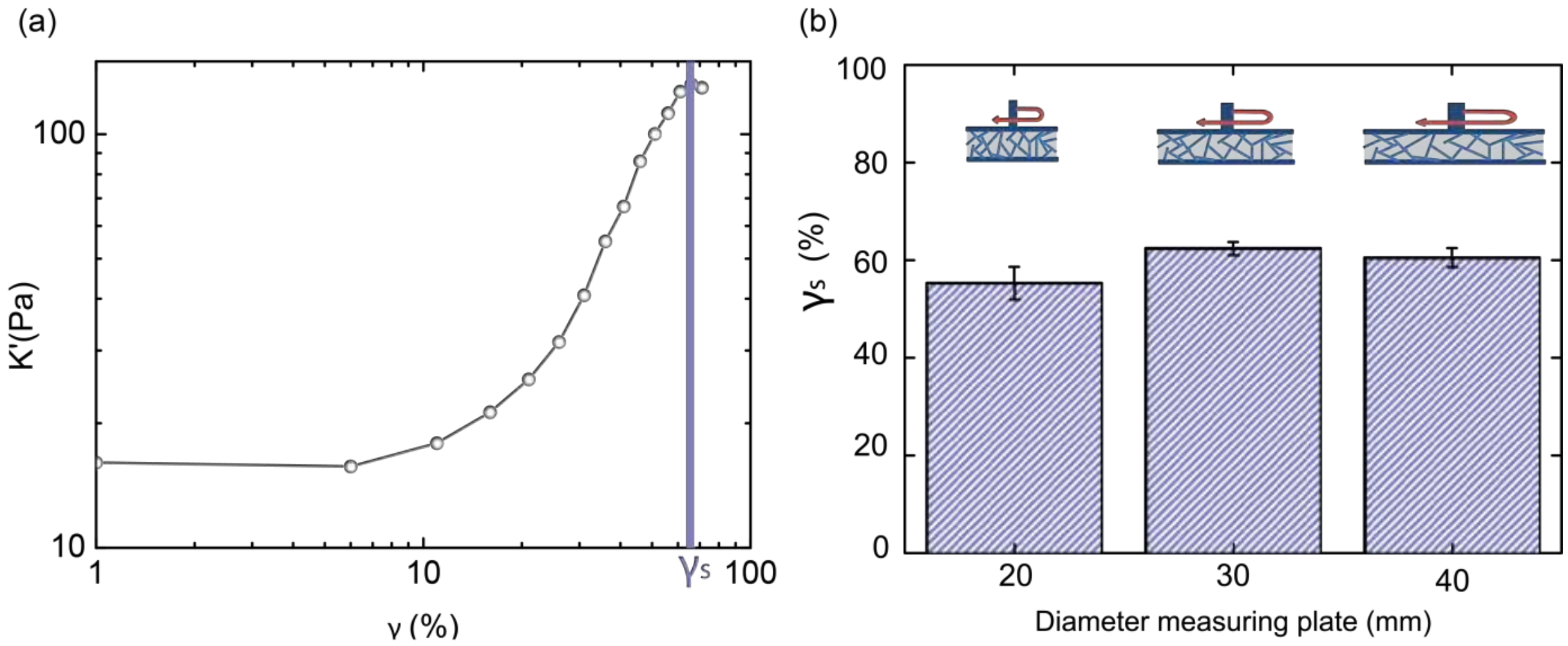

**Supplementary Figure S6: The softening strain is independent of the diameter of the cone-plate measuring geometry. on the softening strain as measured with a stress-ramp (0.1 Pa/s).** (a) Determination of softening strain from a stress-ramp. The softening strain is defined as the strain where the differential elastic modulus *K*' is maximal. (b) Average softening strain for three different rheometer plate diameters (see legend). Data are averages of three independent measurements ± S.E.M.

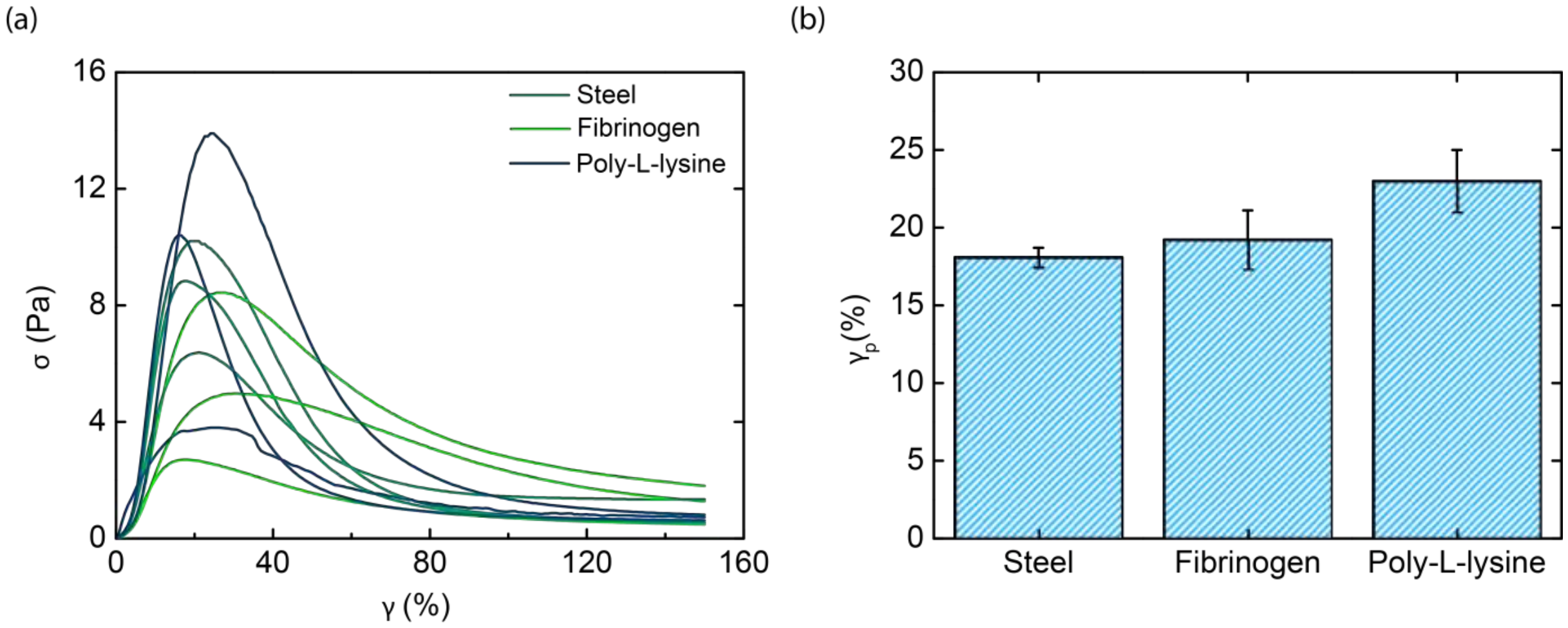

**Supplementary Figure S7: The peak strain of collagen networks is roughly independent of the surface chemistry of the rheometer plates.** (a) Stress-strain curves of 1 mg/ml atelocollagen networks in strain ramps measured with uncoated steel plates (petrol) and for plates coated with either an adhesive fibrinogen-coating (green) or a poly-L-lysine coating (blue). While the peak strain is consistent, the peak stress varies because of intrinsic sample-to-sample variations in the elastic modulus of the networks. Each curve represents a single experiment. (b) Average peak strain for the different surface chemistries, shown as the average of three independent measurements ± S.E.M.

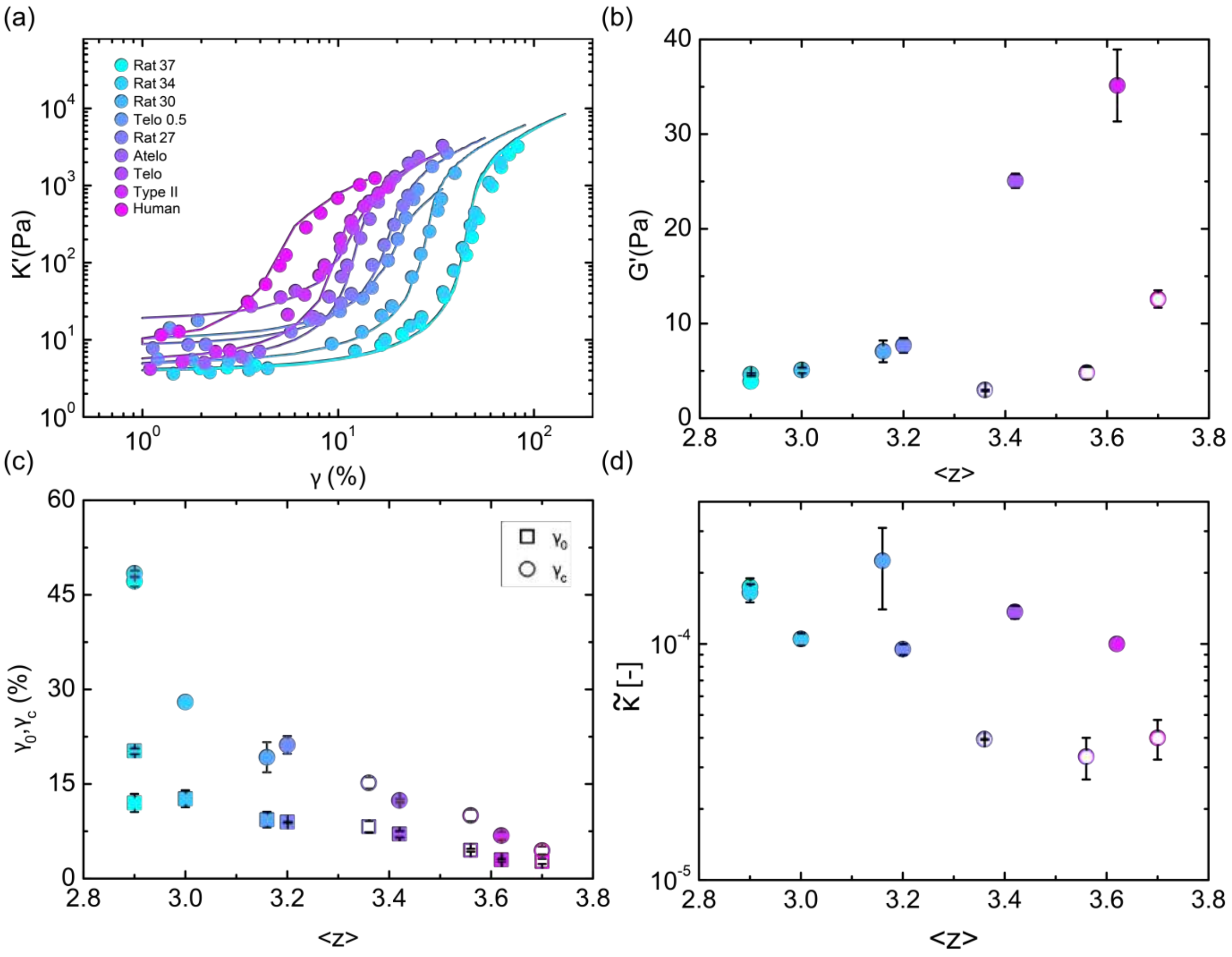

**Supplementary Figure S8: Determination of the connectivity and dimensionless bending rigidity of the collagen networks by calibrating rheology data with simulations of disordered fiber networks.** (a) Measured strain-stiffening curves (symbols) for the different collagen networks (legend) plotted together with fits (solid lines) to Eq. (1) in the main text. (b) The linear elastic modulus of the different collagen networks plotted as a function of connectivity. Solid symbols denote telocollagen (crosslinked) networks and open symbols denote atelocollagen (uncrosslinked) networks. (c) Onset strain $\gamma_0$ for strain-stiffening and critical strain $\gamma_c$ for the transition from soft (bend-dominated) to stiff (stretch-dominated) obtained from the curves shown in panel (a). These data were calibrated against 2D network simulations reported in Ref.[1] in order to infer the average connectivity values <*z*> shown on the x-axis. (d) Dimensionless bending rigidity $\tilde{\kappa}$ of the collagen networks obtained from the fits shown in panel (a). Note that the telocollagens that possess intrafibrillar crosslinking (solid symbols) have a systematically larger $\tilde{\kappa}$ than the atelocollagens that lack intrafibrillar crosslinking (open symbols), indicating that crosslinking makes the collagen fibers behave as more rigid and tight bundles of collagen molecules. The color code based on connectivity is the same as in the main text. Note that in the fitting curves we do not report telocollagen at 30°C, as we cannot reliably fit the stiffening curve.

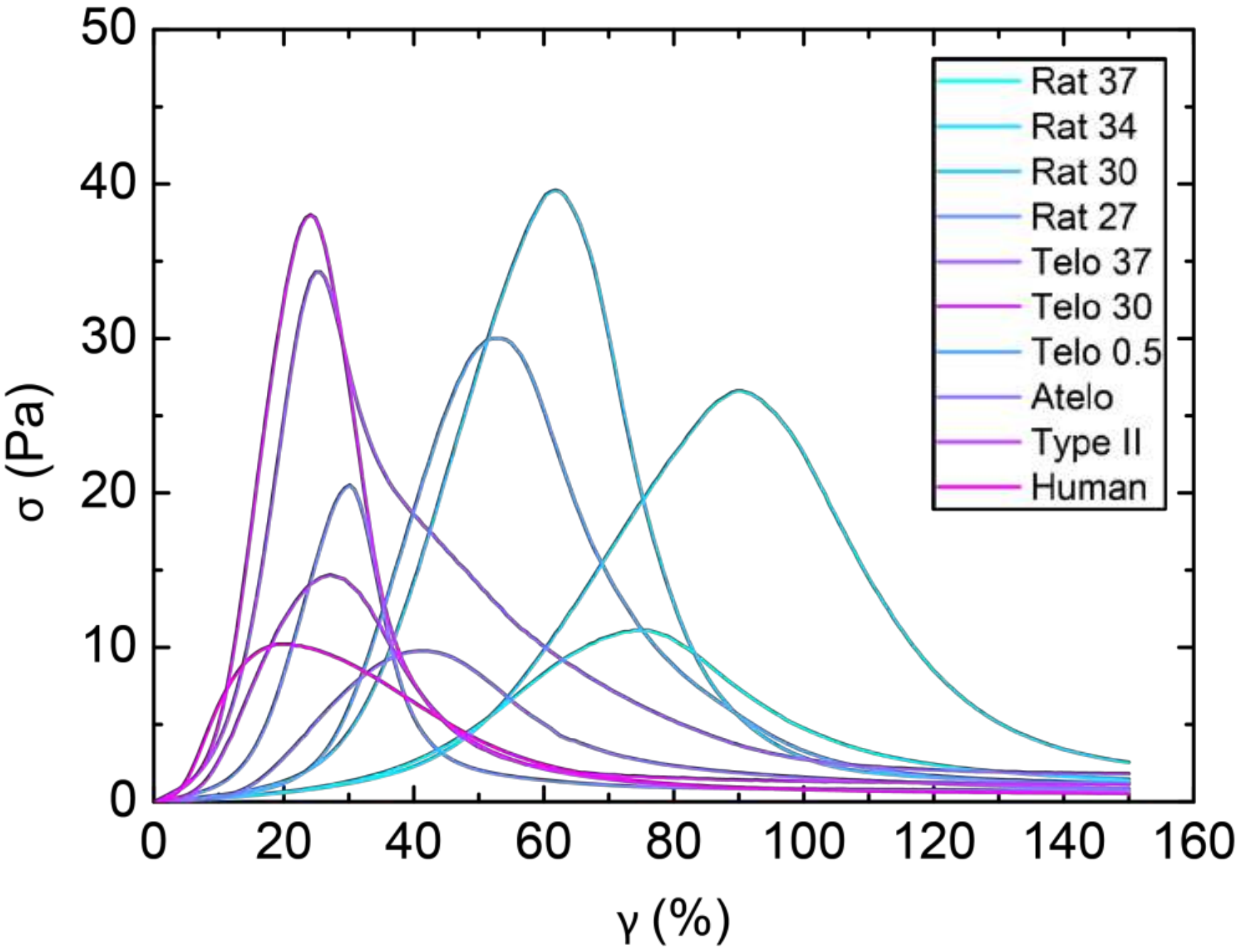

**Supplementary Figure S9: Example stress-strain curves in response to an applied strain ramp for the entire range of collagen networks.** Each curve represents a single experiment. The color code represents the different collagen networks (legend) in the order of increasing connectivity <*z*>.

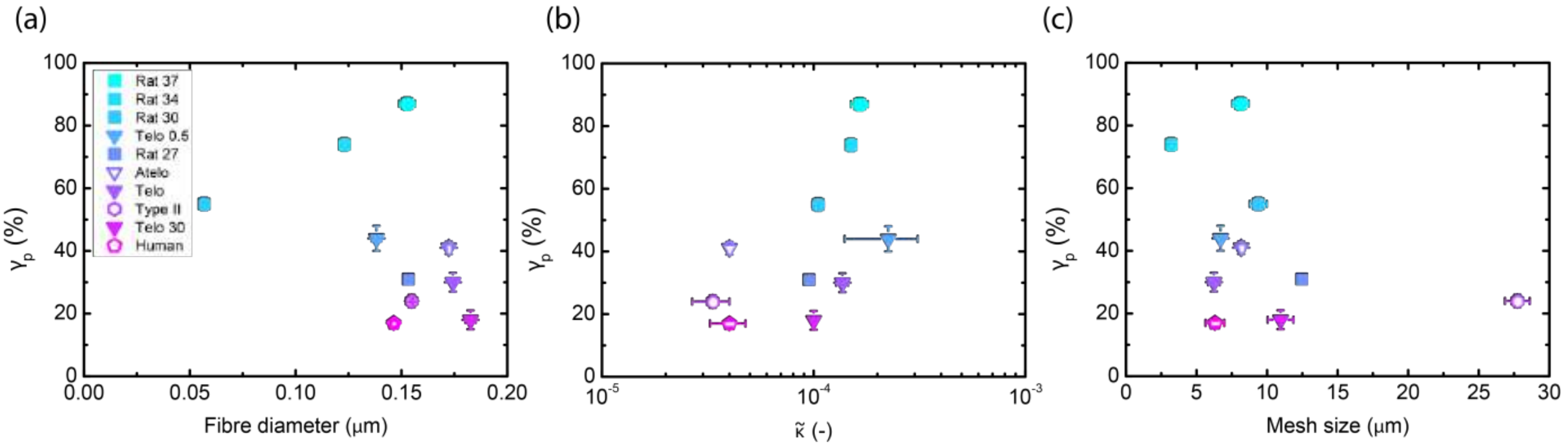

**Supplementary Figure S10: The peak strain shows no correlation with the fiber diameter or bending rigidity nor with the network mesh size.** (a) Scatter plot of peak strain (averaged over at least 3 measurements) against fiber diameter (averaged over at least 100 measurements of fiber diameters in electron microscopy images). (b) Scatter plot of peak strain against dimensionless bending rigidity (averaged over rheology data of at least three different samples). (c) Scatter plot of peak strain against network mesh size (averaged over confocal data from at least three different samples). In all panels, error bars represent the standard error of the mean.

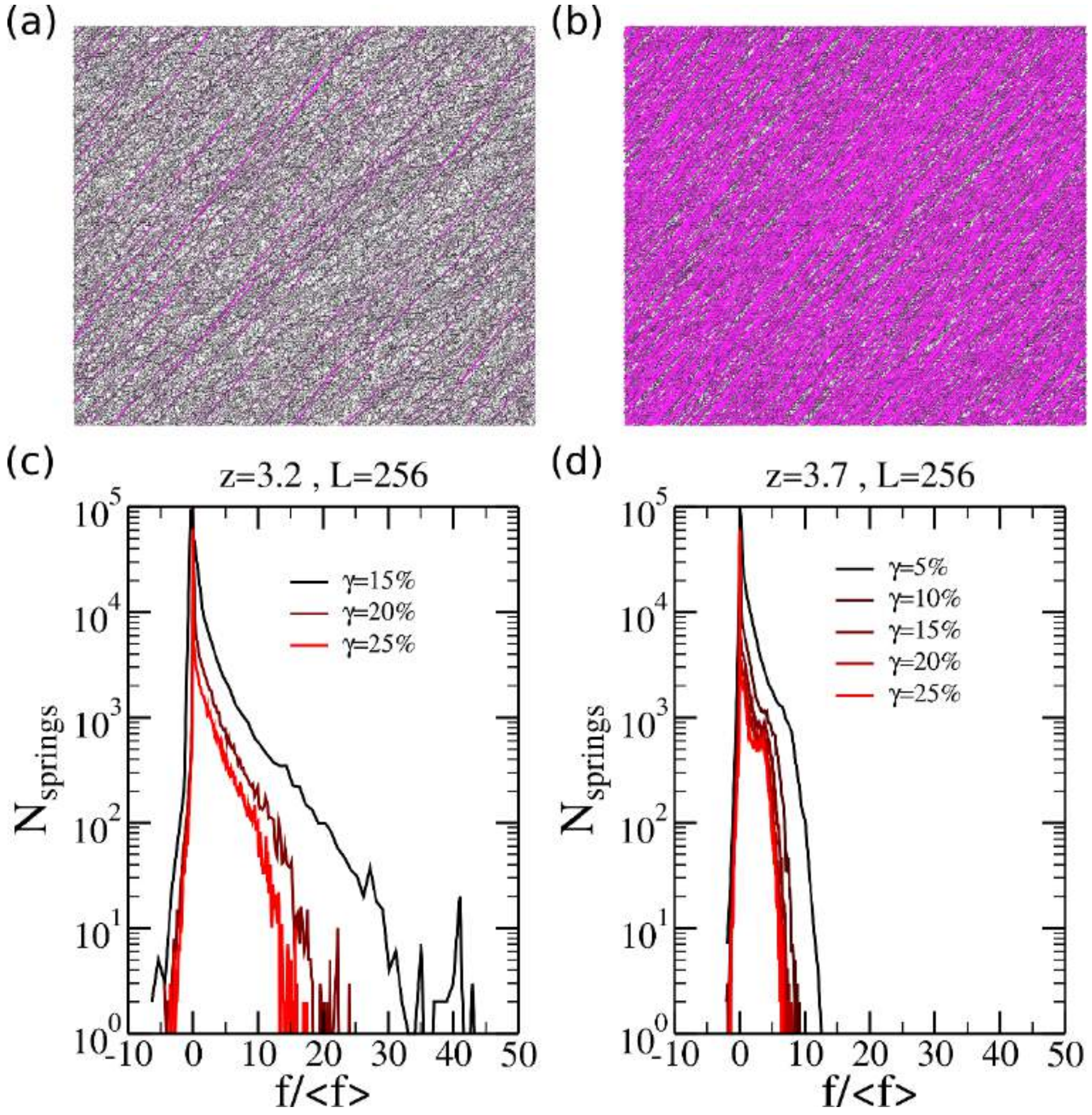

**Supplementary Figure S11: Heterogeneous stress distribution in simulated networks.** Top: Snapshots of networks composed of unbreakable springs under shear deformation with average connectivity (a) <z>=3.2 and (b) <z>=3.7 at a shear strain $\gamma$=20%. Both the color of each spring and the thickness indicate its axial strain, or equivalently the tension it carries (colors range from black for low tension to pink for high tension). Bottom: Histograms of the number of springs carrying a (normalized) force f/<f>, where <f> is the instantaneous average force, for networks with (c) <z>=3.2 and (d) <z>=3.7, in both cases shown for different levels of shear strain $\gamma$ (see legends). The distribution for the more sparsely connected network (<z>=3.2) has a longer tail than for the more highly connected network (<z>=3.7), indicating a more heterogeneous stress distribution (i.e., fewer bonds carry larger forces). The networks have sizes of L×L where L=256.

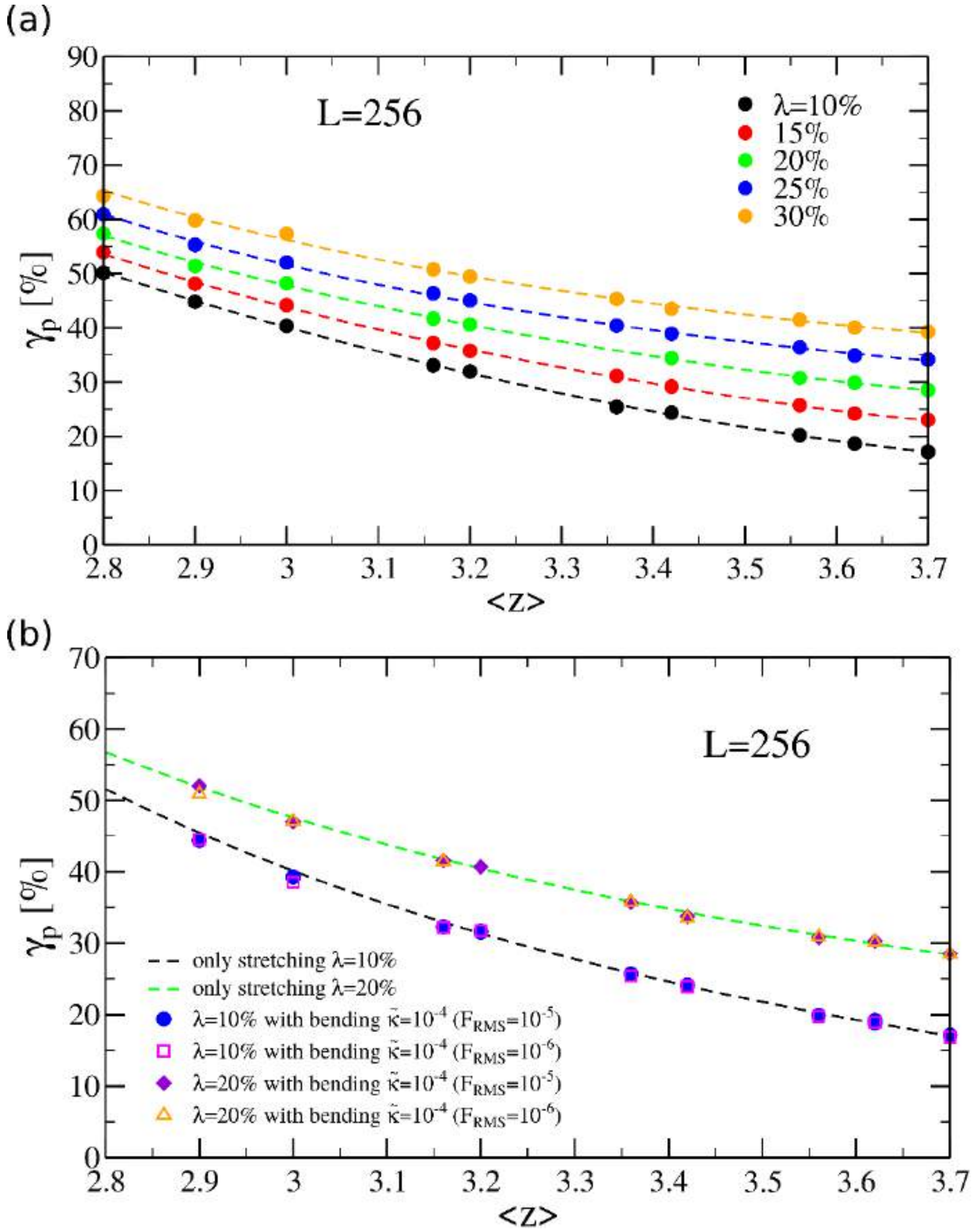

**Supplementary Figure S12: Dependence of the peak strain on the average network connectivity and on the fiber bending stiffness according to simulations.** (a) Results from simulations of large networks (L=256) of springs with zero bending stiffness, performed for different spring rupture thresholds λ between 10 and 30%. Dashed lines are to guide the eye. (b) Results from corresponding simulations of networks with a finite fiber bending stiffness ($\tilde{\kappa}$ =$10^{-4}$). Simulations were performed with different tolerance $F_{rms}$ for the energy minimization procedure ($F_{rms}$ =$10^{-5}$-$10^{-6}$). Results with different $F_{rms}$ are equivalent, therefore confirming that simulations are correctly performed in the athermal limit. The results (symbols) are equivalent to the simulations for springs that only stretch (dashed lines taken from panel (a)), demonstrating that the fiber bending stiffness does not impact the network peak strain for the range of λ and $\tilde{\kappa}$ considered in this study.

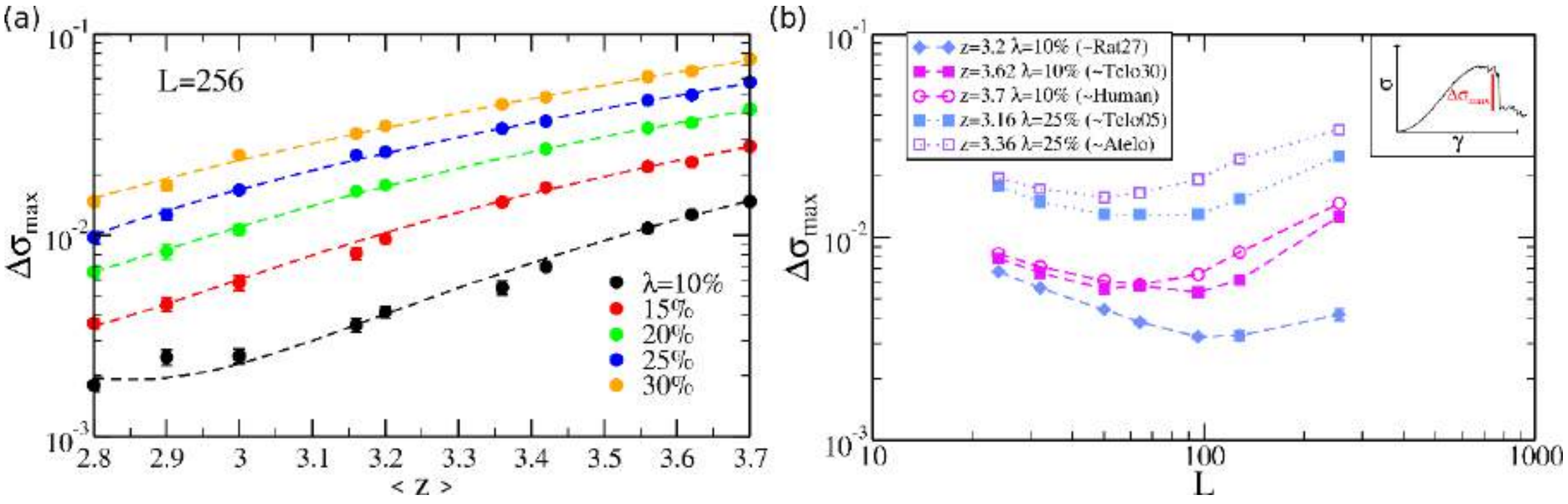

**Supplementary Figure S13: Simulations reveal size-induced brittleness in disordered spring networks.** (a) Brittleness defined as the abruptness of shear-induced fracture, quantified through the maximum stress drop $\Delta\sigma_{max}$ in the same way as in Ref. [2]. See also inset of panel (b). Simulations were performed for networks with different connectivities $<z>$= (x-axis) and different values of the spring rupture threshold λ (see legend), but a fixed system size L=256. (b) System size dependence of $\Delta\sigma_{max}$ for networks with different parameters z and λ that are chosen to approximately match the parameters of several selected collagen networks studied experimentally (see legend). Similar to the fracture response of spring networks subject to uniaxial stretch [2], networks under shear show size-induced brittleness, indicating that this effect is universal with respect to the deformation mode. Dashed lines in both panels are to guide the eye.

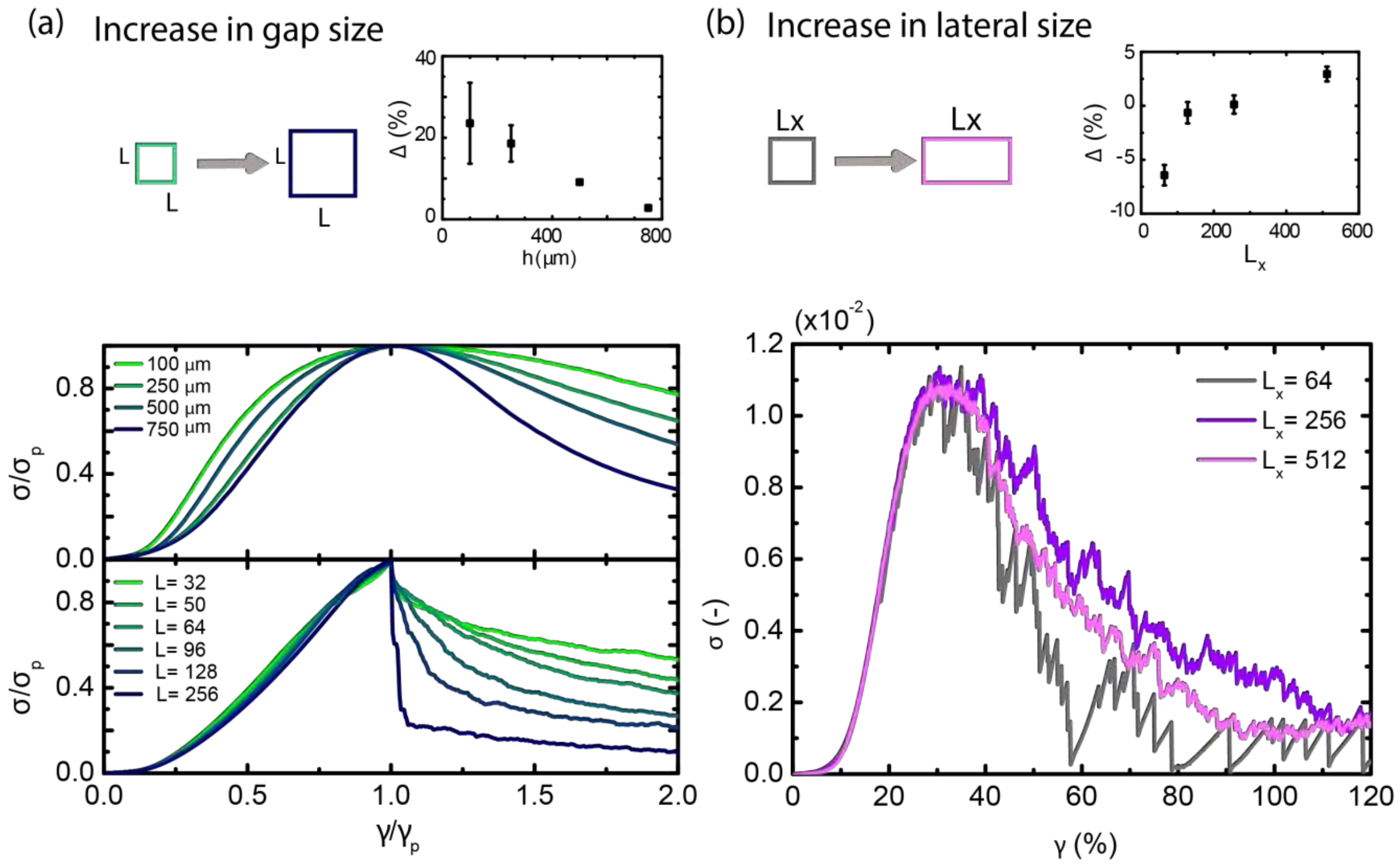

**Supplementary Figure S14: Fracture brittleness is masked by large lateral dimensions in the experiments.** (a) *Upper panel:* The asymmetry between the pre- and post-peak response, quantified by $\Delta$, decreases with gap size, indicating a more brittle response for larger $h$. We define the asymmetry parameter $\Delta=(\gamma^{II}-\gamma_p)-(\gamma_p-\gamma^{I})$, with $\gamma^{I}<\gamma_p<\gamma^{II}$ and where $\gamma^{I}$ and $\gamma^{II}$ are the strains at which the stress is 70% of the peak stress, i.e. $\sigma(\gamma^{I})=\sigma(\gamma^{II})=0.7\sigma_p$. Note that similar results are obtained with different choices of the arbitrary threshold value of $0.7\sigma_p$. $\Delta=0$ corresponds to a perfectly symmetric curve. $\Delta$ can assume both positive and negative values. The smaller the value of $\Delta$, the more brittle the fracture. *Middle panel:* experimental stress-strain curves (human atelocollagen, polymerized at 37°C) measured for different gap sizes $h$, from 100 μm (blue) to 750 μm (green). Each curve was normalized by the peak strain $\gamma_p$ and peak stress $\sigma_p$ and averaged over three samples. *Lower panel*: in simulations of networks composed of $L$ by $L$ nodes (see legend), it is evident that brittleness increases for larger $L$ ($<z>=3.7$, $\lambda=10\%$). (b) A more ductile ("smooth") post-peak response is observed also in simulations when the lateral dimension $L_x$ is increased while keeping $L_y$ (the gap size) fixed, approaching the experimental conditions. *Upper panel:* asymmetry parameter increases upon increasing $L_x$, indicating that fracture becomes more ductile. *Lower panel* shows representative stress-strain curves for a single network with parameters indicated in the legend. Note that stress drops become much smaller upon increasing $L_x$.

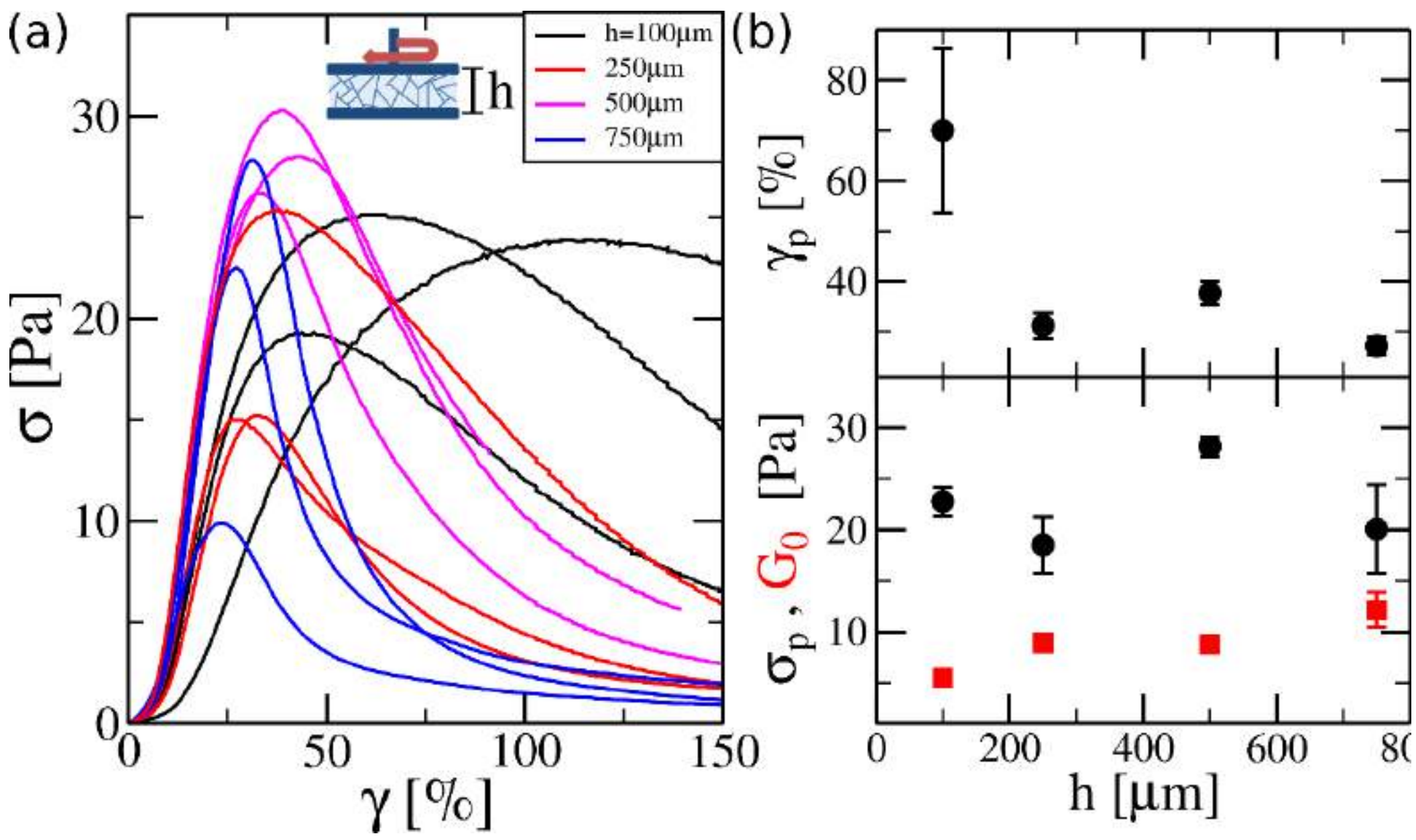

**Supplementary Figure S15: Effect of gap size on fracture of collagen networks (experiments).** (a) Individual stress-strain curves for networks of human atelocollagen measured in a plate-plate geometry with different gap sizes *h*, ranging from 100 to 750 μm (see legend). (b) Peak strain (top) and the linear elastic modulus (bottom, red squares) and peak stress (bottom, black circles) of the networks as a function of *h*. For gap sizes above 200 μm, no significant differences are observed, although the shape of the stress-strain curves still do change with *h*. Data are averages over three independent measurements ± S.E.M.

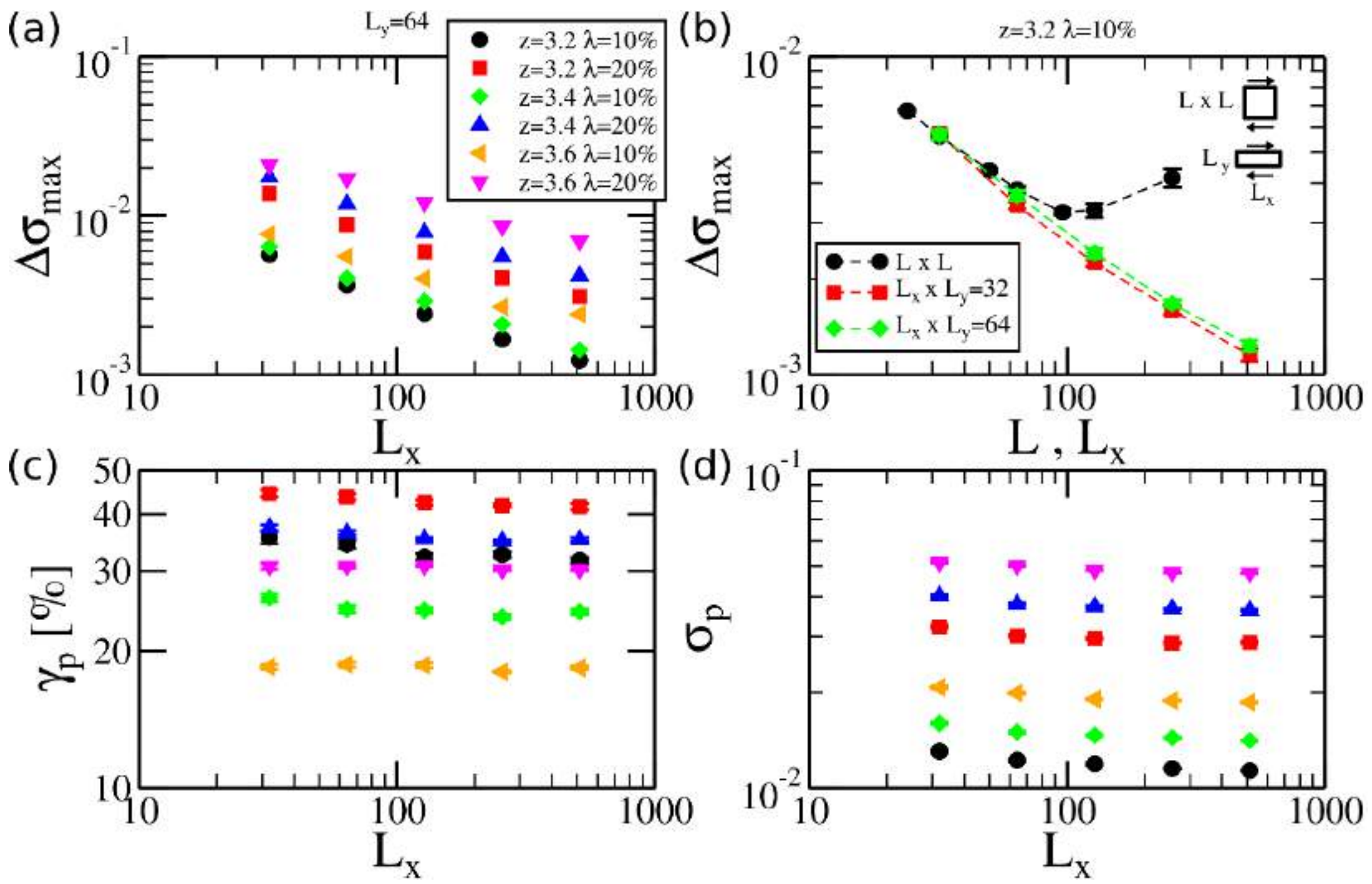

**Supplementary Figure S16: Effect of (lateral) system size on brittleness in simulations.** (a) Maximum stress drop $\Delta\sigma_{max}$ (see Fig. S13 for schematic definition) as a function of lateral system size $L_x$ with fixed $L_y$=64. Fracture is more ductile with increasing $L_x$, as evidenced by a reduction of $\Delta\sigma_{max}$. (b) Size-scaling of $\Delta\sigma_{max}$ for square networks of L×L nodes with varying L (black symbols) and for rectangular networks with $L_y$=32 (red) or $L_y$=64 (green) with varying $L_x$. For the square networks, the stress drop exhibits a minimum, related to the size-induced brittleness studied in our previous work [2]. By contrast, the maximum stress drop for the rectangular networks with fixed $L_y$ monotonically decreases with increasing $L_x$. Lines are to guide the eye. (c-d) No significant size-dependence of the peak strain (c) and peak stress (d) is observed for the rectangular networks with fixed gap when only the lateral dimension is increased, differently from the square L×L networks (see Supplementary Figure S17). The different symbol colors and shapes in all panels correspond to different values of the average network connectivity <*z*> and spring rupture threshold *λ* (see legends).

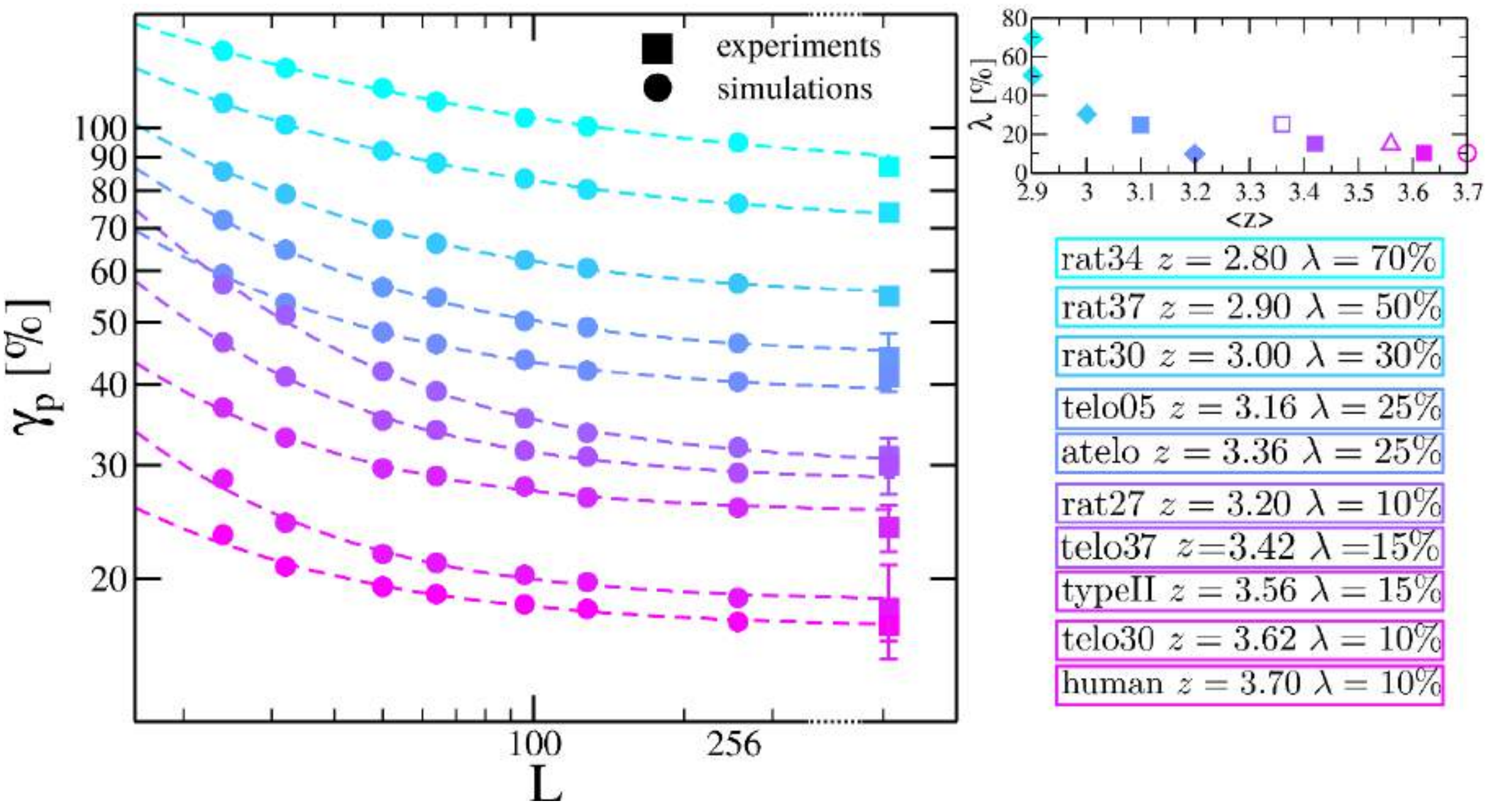

**Supplementary Figure S17: Identification of effective fiber rupture threshold by mapping experiments on simulations, while accounting for size-scaling.** Left: Size dependence of the peak strain for simulations of L×L spring networks with average connectivity <$z$> and rupture threshold λ chosen to match the experimental values of $\gamma_p$ in the large system size limit. Dashed lines are best power-law fists used for the extrapolation to large L. Right: (effective) fiber rupture threshold for the whole range of collagen networks as a function of their average connectivity. The fiber rupture threshold on the y-axis is obtained by mapping the measured network peak strain $\gamma_p$ onto simulations, while the connectivity values on the x-axis are obtained by mapping the strain-stiffening curves onto simulations (Figure S8a). Open symbols indicate atelocollagen (uncrosslinked) networks, while closed symbols indicate telocollagen (crosslinked) networks. The data are arranged in order of increasing $z$ (cyan for < $z$>=2.9 to purple for < $z$>=3.7). At lower <$z$> a significantly higher fiber rupture threshold λ is predicted based on the network model with no spring lengthening.

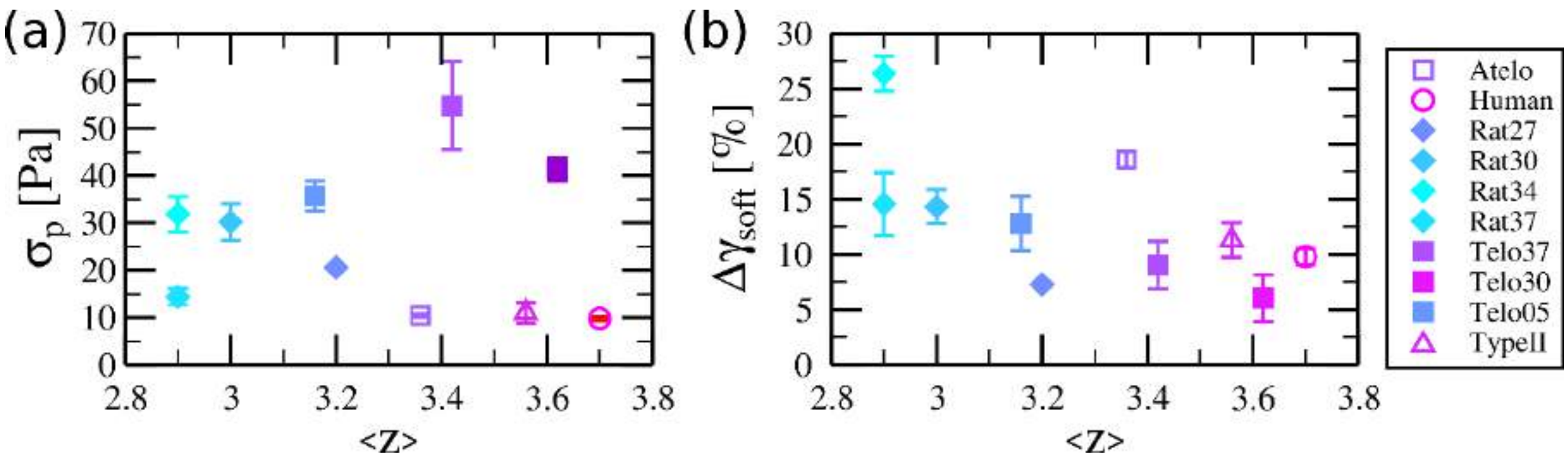

**Supplementary Figure S18: Atelocollagen (uncrosslinked) networks are softer than telocollagen (crosslinked) networks and exhibit a larger softening interval (experiments).** (a) Peak stress for different collagen networks as a function of their average connectivity. (b) Corresponding softening interval, defined as the strain interval between the strain where the differential elastic modulus is maximal and the peak strain $\gamma_p$ where the modulus is zero. Note that, for the same connectivity, the atecollagen (uncrosslinked) networks (open symbols) tend to exhibit a smaller peak stress $\sigma_p$ and larger softening interval than the telocollagen (crosslinked) networks (solid symbols), consistent with uncrosslinked collagen fibers being softer and more plastic than crosslinked fibers.

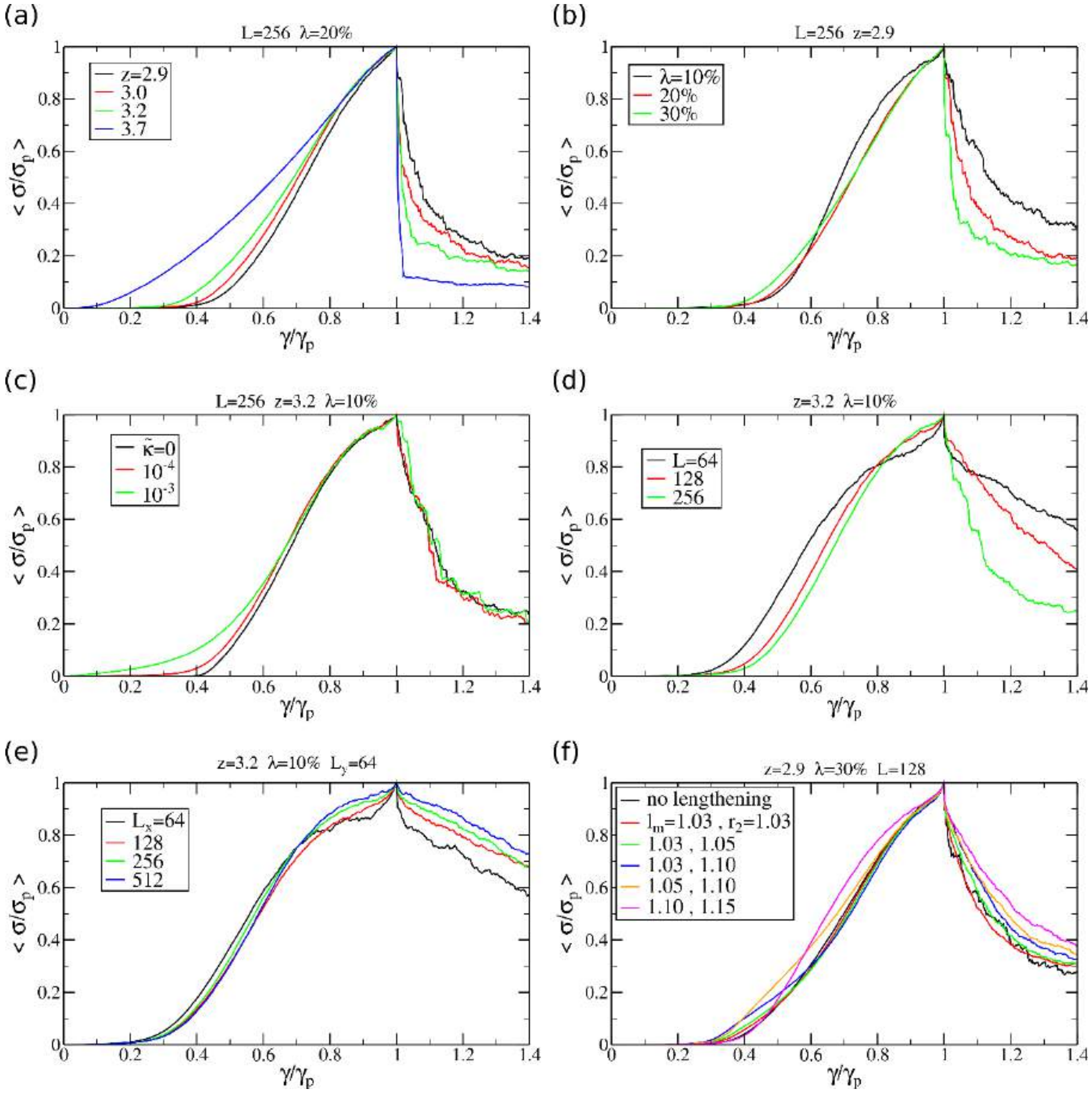

**Supplementary Figure S19: Dependence of the shape of the stress-strain response in simulations on the simulation parameters.** (a) Effect of connectivity <z> (see legend) on the normalized and averaged stress-strain curves at fixed system size L=256 and rupture threshold λ=20%. (b) Corresponding simulation results showing the effect of varying rupture threshold λ for fixed system size L=256 and connectivity <z>=2.9. (c) Corresponding data for varying dimensionless bending rigidity $\tilde{\kappa}$ (with fixed L=256, λ=10%, <z>=3.2) reveal that including a finite bending rigidity only shifts the initial part of the stress-strain curve (the linear regime) upwards with respect to networks of springs with $\tilde{\kappa}$=0, while the behavior close to rupture is $\tilde{\kappa}$-independent. (d) Effect of system size for L×L networks on the stress-strain response, for fixed <z>=3.2 and λ=10%. (e) Corresponding data for $L_x \times L_y$ networks with varying $L_x$ and fixed $L_y$=64, <z>=3.2, λ=10%. (f) Effect of plastic effects modelled as an irreversible increase in spring rest length from $r_1$=1.0 to different rest lengths $r_2$ (between 1.03 – 1.15) at different onset lengths $l_m$ (between 1.03 – 1.10) compared to the elastic limit of springs that do not lengthen (black curve).

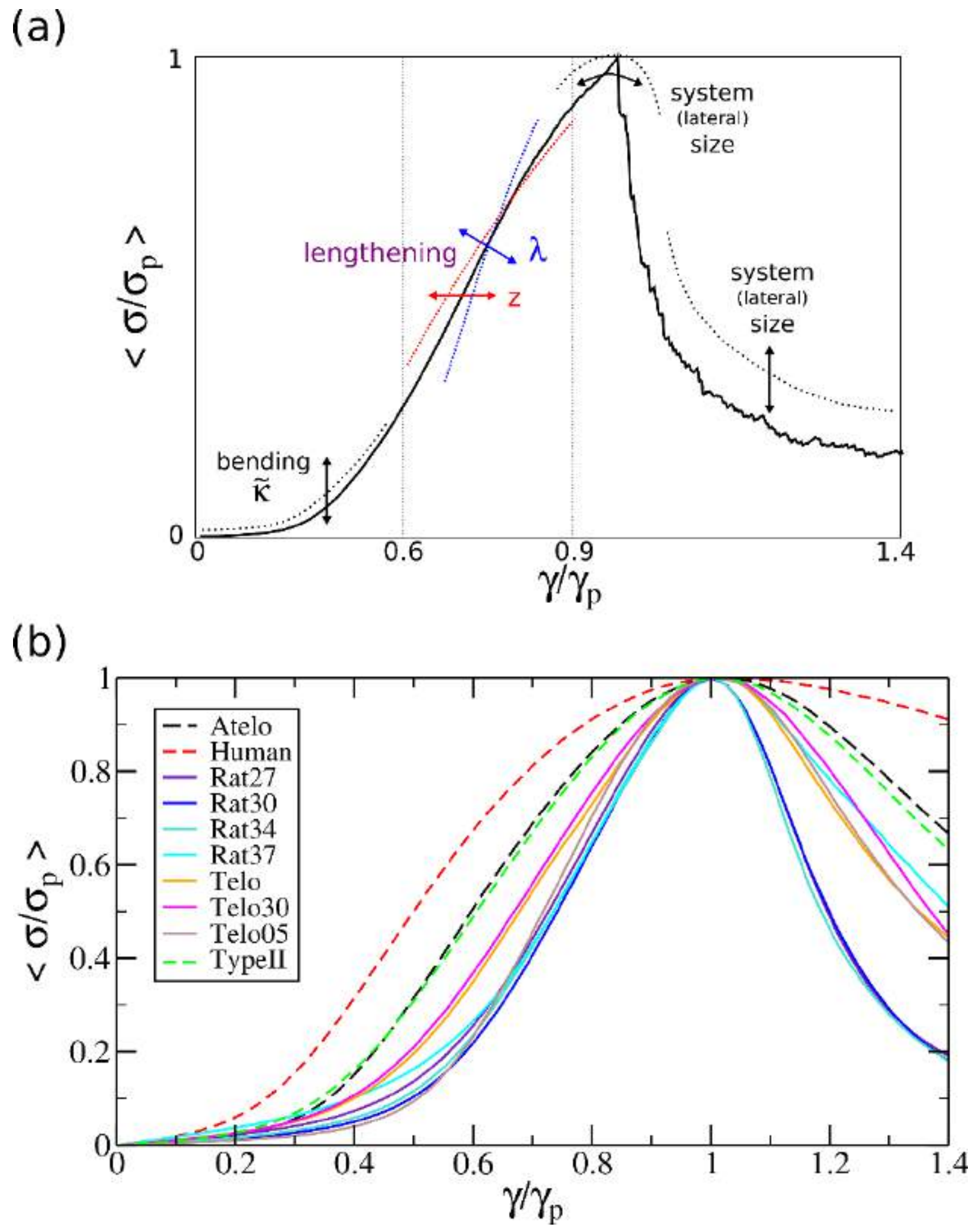

**Supplementary Figure S20: The shape of the stress-strain response of collagen networks in simulations and experiments reveals differences in plasticity depending on network connectivity and intrafibrillar crosslinking.** (a) Schematic summarizing the influence of the different parameters in the simulations on the shape of the normalized stress-strain curves. We match the middle region of the curves (i.e., in the reduced strain interval $0.6 < \gamma/\gamma_p < 0.9$, indicated by the vertical dashed lines) measured experimentally against simulation results in order to assess the role of fiber lengthening in the network response. (b) Experimental normalized and averaged curves for all the collagen networks investigated in this study. Note that the curves associated with atelocollagen (uncrosslinked) networks (Atelo, Human, TypeII, shown with dashed lines) have a different shape from the other curves obtained for telocollagen (crosslinked) networks, indicating enhanced fiber plasticity.

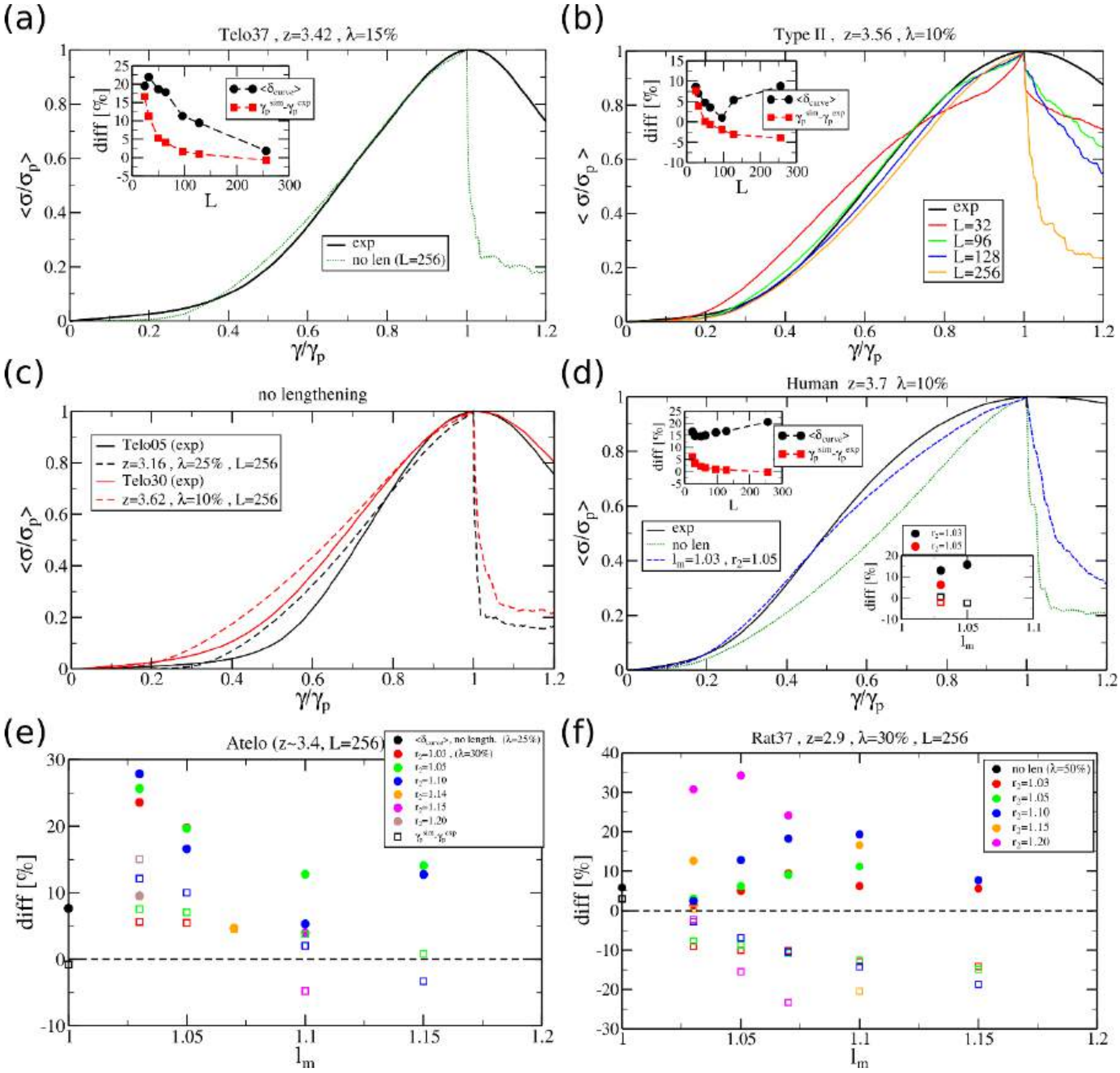

**Supplementary Figure S21: Plasticity in collagen networks inferred from matching experiments with simulations performed for different fiber rupture and lengthening parameters.** (a) Comparison of normalized and averaged stress-strain curve from experiments on telocollagen (crosslinked) networks (black solid line) with simulations for large networks (L=256, <z>=3.42) of springs that do not lengthen (λ=15%). Inset: system size dependence of the average relative difference $<\delta_{curve}>$ between the simulation and experimental curves in the interval $0.6 < \gamma/\gamma_p < 0.9$ (black circles) and of the difference in peak strain $\gamma^{sim} - \gamma^{exp}$ (red squares). The best match is obtained for large system sizes, consistent with the large rheometer gap size (250μm) compared to the network mesh size (~5μm). (b) Same comparison for type II collagen networks. In this case the best match is obtained for intermediate system size of the simulation box (L~96), consistent with the >3-fold larger mesh size (~27μm) of the type II networks compared to the collagen I networks. (c) Same comparison for the other two telocollagen (crosslinked) networks (in both cases from bovine dermis), showing that the trend with changing connectivity (<*z*>=3.16 versus <*z*>=3.62) is captured by the simulations. (d) Comparison for the human atelocollagen (uncrosslinked) network (solid black line) with simulations in the elastic limit (green dotted line) and for springs that lengthen (blue dashed line). In this case, including plastic effects in the simulation is necessary to match the experimental response. Note that since λ=10% is small and close to $l_m$=1.03, we cannot distinguish between plasticity stemming from

the fiber lengthening versus bundle/branch opening. Left inset: system size dependence of <$\delta_{curve}$> (black circles) and $\gamma^{sim}$ - $\gamma^{exp}$ (red squares) as in panels (a,b). Right inset: <$\delta_{curve}$> for two different values of $r_2$. (e) Differences <$\delta_{curve}$> between experimental curves for atelocollagen (uncrosslinked networks) and simulation curves (solid circles) and between the corresponding experimental and simulation peak strains (open squares), for various combinations of the fiber lengthening parameters $r_2$ and $l_m$. Good agreement is obtained when $l_m \approx 1.10$. (f) Same as panel (e), for Rat37 telocollagen (crosslinked) networks. In this case, better agreement is obtained for small $l_m$, suggesting plasticity at the network level.

## REFERENCES SUPPLEMENTARY INFORMATION

[1] Jansen, K. A. *et al*. The Role of Network Architecture in Collagen Mechanics. *Biophys. J.* **114**, 2665–2678 (2018)

[2] Dussi, S., Tauber, J. & van der Gucht, J. Athermal Fracture of Elastic Networks: How Rigidity Challenges the Unavoidable Size-Induced Brittleness. 1–13 (2019). arXiv:1907.11466

## Supplementary Video

**Supplementary Video S1:** Video corresponding to the fracture process of the collagen network shown in Figure 1. Frame rate is 3.5 fps. Scale bar indicates 10 μm.